\documentclass[amsmath,amssymb,showpacs,showkeywords,twocolumn]{revtex4}
\usepackage[dvips]{graphicx,color}
\usepackage{times}
\usepackage{mathrsfs}
\usepackage{amsmath}
\usepackage{graphicx}
\usepackage{epsfig}
\usepackage{dcolumn}
\usepackage{bm}
\begin{document}
\title{Signature of Anomalous Andreev bound states in magnetic Josephson junction of noncentrosymmetric superconductor on a topological insulator}
\author{Saumen Acharjee\footnote{saumenacharjee@gmail.com} and Umananda Dev 
Goswami\footnote{umananda2@gmail.com}}
\affiliation{Department of Physics, Dibrugarh University, Dibrugarh 786 004, 
Assam, India,}

\begin{abstract}
We study the Josephson effect in a clean noncentrosymmetric superconductor/half-metal/noncentrosymmetric superconductor junction, which is grown on the 
surface of a three-dimensional Topological Insulator (TI) in the ballistic 
limit. We find the signature of anomalous Andreev Bound States (ABS) and band 
splitting for a spin-active barrier whose barrier magnetic moment is misaligned
with the bulk moment. The chiral Majorana mode and 4$\pi$ periodic ABS are 
found to exist on the surface of TI for parallel orientation of the moments 
in the normal incidence condition. But for anti-parallel misalignment, we 
observe the 2$\pi$ periodic ABS. There exist a gap in ABS for oblique 
incidence. We find the splitting of Andreev levels in the presence of 
RSOC and also for unequal mixing of singlet-triplet correlations present in 
NCSC. The Majorana mode, ABS and Josephson supercurrent can be controlled by 
the ratio of barrier magnetic and non-magnetic moments. The critical current 
is found to be maximum for singlet or triplet dominated NCSC, while it is 
minimum in the equal mixing condition. The ABS is found to be barrier 
thickness dependent and is suppressed for an opaque barrier. We observe a 
monotonic decay in critical current with a finite length of the junction for 
all the singlet-triplet mixings and magnetic moments. The current-phase 
relation is found to be sinusoidal with no phase shift in the half-metallic 
limit, however, for different orientations of the bulk moment, an anomalous 
characteristic is also observed.

\end{abstract}

\maketitle

\section{Introduction}
Topological Insulators (TI) are a new class of material that exhibit 
non-trivial band topology due to a strong spin-orbit coupling \cite{fu1,fu2,
moore2,roy,moore1,hasan1,qi,hasan2}. They remain insulating in bulk, but the 
most intriguing characteristic of them is the presence of gapless conducting 
edge states (in two dimensions) and surface states (in three dimensions). In 3D
TIs, the spin of a moving particle is helically locked to the momentum at 
their surface \cite{beidenkopf,roushan,liu,dyrdal,li,yang,chen12}. Due to 
spin-momentum locking, it is possible to induce new phases when a magnetic 
or superconducting element is brought in close proximity with the 3D TIs.
Previous works indicate that when a conventional s-wave superconductor is 
coupled with 3D TIs then the superconducting correlation is induced at their 
surface \cite {fu3,fu4,sochnikov,kohda}. In contrast to the
conventional spin-singlet Cooper pairing, the induced superconductivity is an 
admixture of both singlet and triplet correlations due to spin lifted 
degeneracy at their surface \cite{gorkov1,tkachov, gong}. Moreover, the 
Andreev Bound States (ABS) formed in superconductor-topological 
insulator-superconductor (S$|$TI$|$S) Josephson junctions can exhibit zero 
energy crossing, when the phase difference between the superconductor is 
$\pi$. Thus it can host Majorana quasiparticle \cite{gorkov1,tkachov,
sochnikov,kohda, gong,fu3,fu4,tanaka1}, which is a fermionic mode of its own 
antiparticle. Though in conventional S$|$N$|$S junctions, a weak backscattering
can splits the spin degeneracy at $\pi$, however, due to non-trivial topology, 
S$|$TI$|$S junctions can support $4\pi$-periodic Josephson supercurrent. 
Recently, their is a growing interest 
to understand the robust phenomena like, zero-energy Majorana modes 
\cite{gorkov1,tkachov,sochnikov,kohda, gong,fu3,fu4,tanaka1} and topological 
superconductivity \cite{frolov,chen11,linder11,hsieh,nakhmedov,trang} on 
the surface of 3D TIs. Experimentally it is very challenging to distinguish 
topologically trivial and non-trivial $4\pi$ modes. It is due to the reason 
that most of the TIs discovered till now are not perfectly insulating in bulk. 
But recently, it is reported that the majority of helical supercurrent 
is due to its surface and only a feeble contribution arises due to the bulk 
\cite{veldhorst,cho,stehno}. Moreover, some exotic TIs are also discovered 
which display insulating bulk at low temperatures \cite{kayyalha,jauregui,
xu11,xu12}. 

Many theoretical and experimental efforts are made to demonstrate helical ABS, 
Current Phase Relation (CPR) and temperature dependence of critical current in 
various S$|$TI$|$S Josephson junctions \cite{kwon,olund,burset,ghaemi,tkachov1,
snelder1,snelder2,kurter,schrade,linder12}. However, the ABS and the 
supercurrent formed on the helical surface states due to Non Centrosymmetric 
Superconductors (NCSC) are yet to be understood. The reasons for considering 
NCSC are: (i) Coexistence of spin-singlet and spin-triplet superconducting 
correlations and (ii) the existence of strong Antisymmetric Spin-Orbit 
Coupling (ASOC) due to the absence of inversion symmetry \cite{bauer1,bauer2,
bauer3,motoyama, kawasaki, akazawa, anand1, smidman, anand2, yuan, togano,
badica, matthias, singh, pecharsky, hillier, bonalde, yogi, ali, xu,acharjee2,
acharjee3,acharjee1,linder15,frigeri}. Earlier studies indicate that due to 
the presence of the Rashba Spin-Orbit Coupling (RSOC) the pairing symmetry and 
hence the transport properties drastically get influenced in NCSC 
heterostructure \cite{yuan, togano, badica,acharjee2,acharjee3,linder15,
frigeri}. Moreover, it is also reported that the RSOC and singlet-triplet 
mixing ratio of NCSC play very significant roles in the formation of ABS and 
supercurrent \cite{vorontsov,klam,zhang11,smidman1,duckheim,sun,iniotakis}.  
So it is necessary to study the interplay of RSOC and singlet-triplet 
correlation in helical ABS and also to understand the Majorana modes in NCSC 
Josephson junctions.  

Josephson junctions with a magnetic material as an intermediate layer can 
exhibit unconventional spin-triplet proximity effect \cite{bobkova, saxena,
aoki,pfleiderer,huy,buzdin,rahnavard,zutic}. As spin-triplet correlations are 
long-ranged \cite{keizer1,eschrig,khaire,linder14}, so the supercurrent in the 
magnetic element of such heterostructures can be carried by spin-triplet 
Cooper pairs. The long-range nature of spin-triplet correlation is due to a 
combination of two mechanisms: (i) spin-flip scattering \cite{eschrig2, 
galaktionov, eschrig} and (ii) spin-active interfacial phase shifts of the 
wave functions of electrons \cite{millis, hubler, galaktionov}. Thus 
the superconductor-ferromagnet heterostructures gained a lot of research 
interest in the present era. Also, with the development of spintronic devices, 
Half Metallic (HM) ferromagnets have attracted a particular attention due to 
high spin polarization \cite{eschrig2, galaktionov, eschrig,millis, hubler}. 
Previously, the dc Josephson effect and its dependence on spin-flip parameters 
have been studied considering a spin active barrier in S$|$HM$|$S Josephson 
junctions \cite{galaktionov}. In the present work, we consider an 
NCSC$|$HM$|$NCSC Josephson junction grown on the surface of 3D TI 
(See Fig.~\ref{fig1}). The goal of 
this work is to understand the magnetic-superconducting correlations, the 
interplay between spin-dependent interfacial properties, RSOC and 
superconducting pairing in the bulk NCSC with regard to (i) helical ABS, (ii) 
Josephson supercurrent and (iii) Majorana modes on the surface of 3D TI.  

The paper is organized as follows: In Section II, we present our theoretical 
model of the proposed system and present an explicit form of the Hamiltonian. 
We calculate the ABS and supercurrent both analytically and numerically in 
Section III. In Section IV, we discuss the numerical as well as some 
approximated analytic results. Finally, we summarize our work with conclusions 
in Section V.

\section{Model and formulation}
\begin{figure}[hbt]
\centerline
\centerline{ 
\includegraphics[scale=0.5]{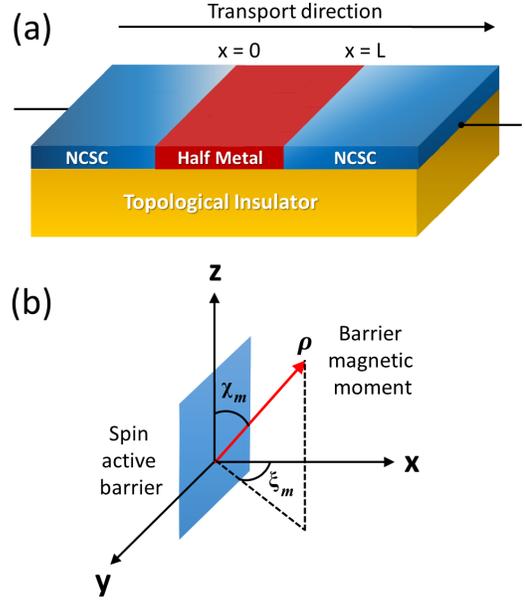}
\vspace{0.1cm}
}
\caption{(a) Schematic illustration of the proposed NCSC$|$HM$|$NCSC Josephson 
junction on the surface of 3D TI. We consider the bound states at the surface 
of a TI which is in strong proximity with the NCSC and a HM. The current is 
supposed to flow on the surface of the TI. (b) Representation of a spin active 
barrier. We consider the bulk magnetization in the HM is along the z-direction 
and the barrier magnetic moment is misaligned with the bulk magnetization by 
angles ($\chi_m$, $\xi_m$).}
\label{fig1}
\end{figure}

We employ Bogoliubov-de Gennes (BdG) formalism to study the bound states and 
Josephson supercurrent (JSC) for the proposed setup as shown in 
Fig.~\ref{fig1}. In Nambu basis
\begin{equation}
\label{eq1}
\Psi = \lbrace{\psi_\uparrow,\psi_{\downarrow},\psi_{\uparrow}^{\dagger},\psi_{\downarrow}^{\dagger}}\rbrace
\end{equation}

\noindent the Hamiltonian for the surface of a 3D TI with induced 
superconducting and magnetic proximity due to NCSC and HM can be defined 
as \cite{snelder1,snelder2,linder12,tanaka1}
\begin{equation}
\label{eq2}
\hat{\mathcal{H}} = \left(
\begin{array}{cc}
 \hat{\mathcal{H}}_0(\mathbf{k}) +\hat{M}  & \hat{\Delta }_{\alpha \beta } \\
 \hat{\Delta }_{\alpha \beta }^{\dagger} & -\hat{\mathcal{H}}^{\dagger}_0(\mathbf{-k}) 
 -\hat{M}^{\dagger} \\
\end{array}\right)
\end{equation}

\noindent where, $\hat{\mathcal{H}}_0(\mathbf{k})$ and $\hat{M}$ are defined as
\begin{equation}
\label{eq3}
\hat{\mathcal{H}}_0(\mathbf{k}) = v_F(\hat{\sigma}_xk_x + \hat{\sigma}_xk_y)-\mu + V_{int},
\end{equation}
\begin{equation}
\label{eq4}
\hat{M} = \mathbf{m}.\boldsymbol{\sigma}\Theta(x)\Theta(L-x)
\end{equation}

\begin{figure}[hbt]
\centerline
\centerline{ 
\includegraphics[scale=0.31]{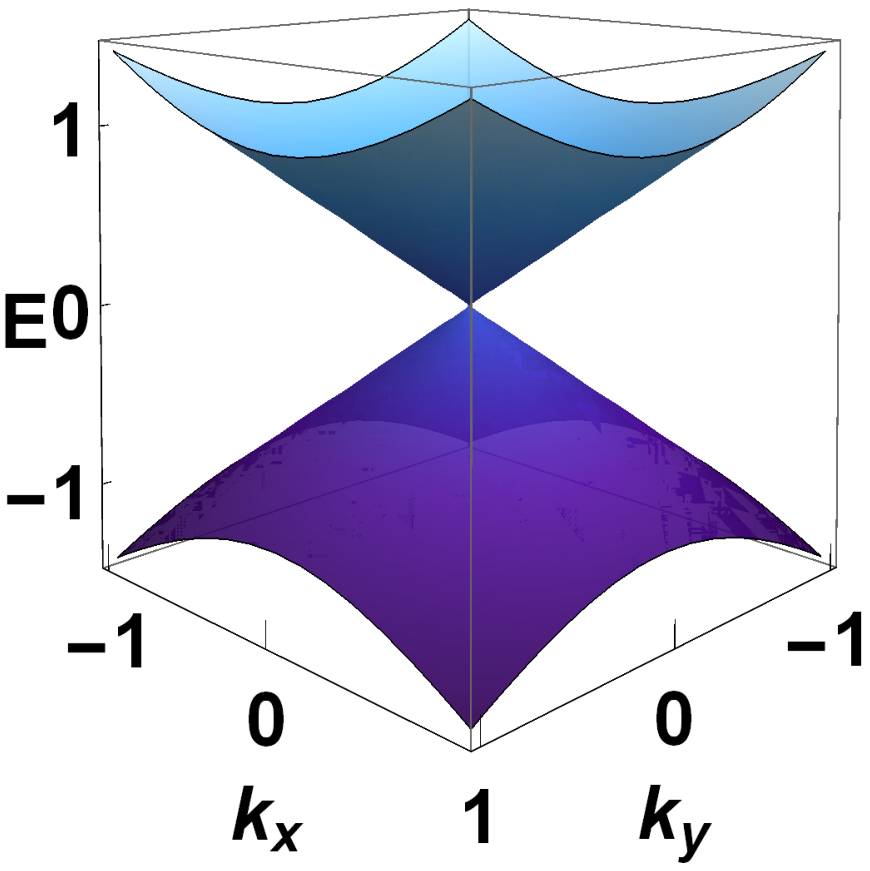}
\vspace{0.1cm}
\includegraphics[scale=0.31]{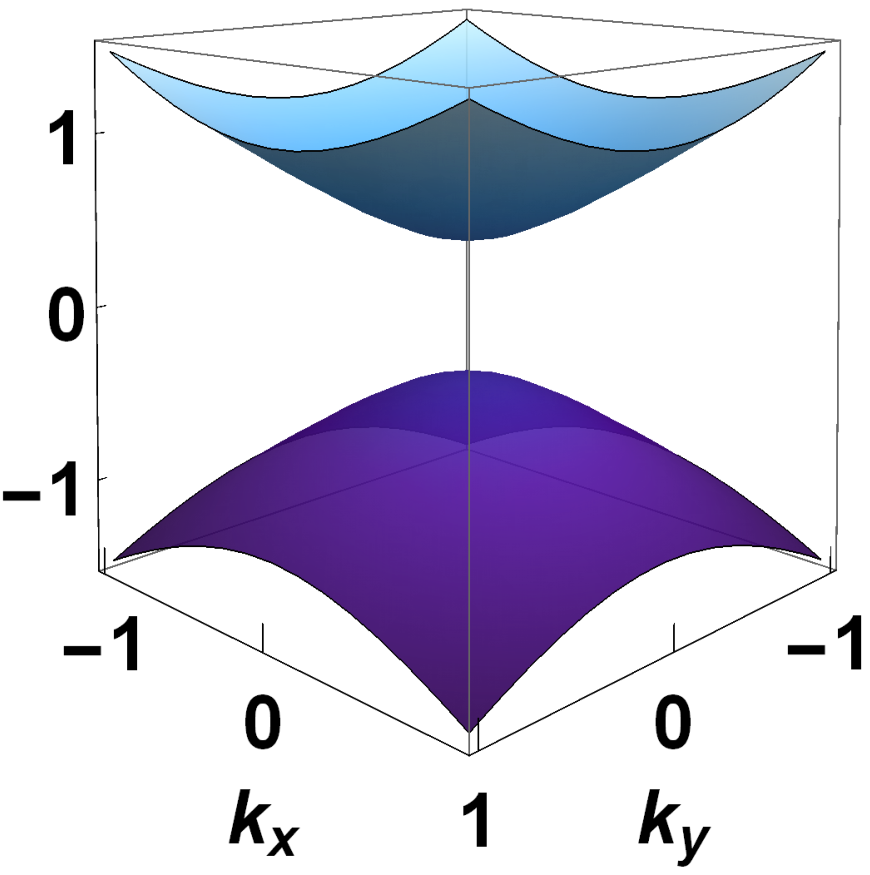}
\vspace{0.1cm}
\includegraphics[scale=0.31]{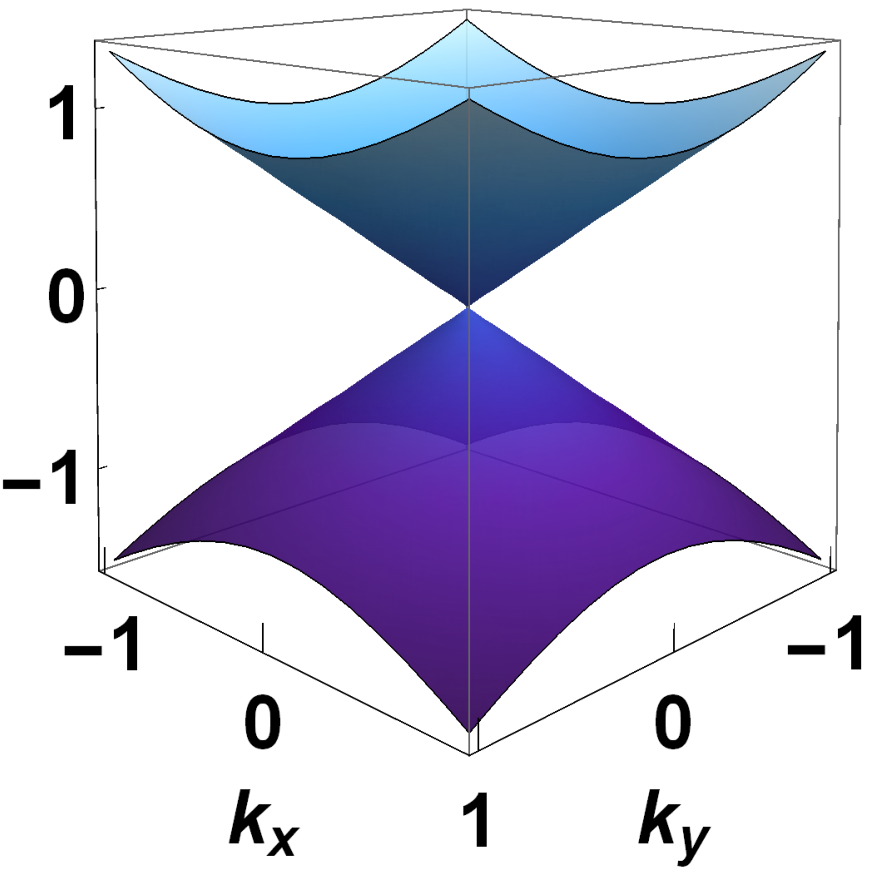}
\vspace{0.1cm}
}
\caption{Energy band spectrum at the surface of TI. The plots in the left and 
middle respectively represent the band spectrum for parallel alignment of 
barrier and bulk magnetization with $\rho = 0$ and $\rho = 0.3$, while the 
plot in the right is for anti-parallel orientation of barrier and bulk 
magnetic moment with $\rho = 0.3$.}
\label{fig2}
\end{figure}
\noindent where, $v_F$ is the Fermi velocity, $\mu = \mu_i[\Theta(-x) + 
\Theta(x-L)]$ are the chemical potentials of the respective layers, 
$\mathbf{m} = (m_x,m_y,m_z)$ is the magnetization of the bulk HM and 
$\boldsymbol{\sigma} =$ ($\sigma_x,\sigma_y,\sigma_z$) are Pauli's spin 
matrices. The last term of Eq.~\eqref{eq3} represents the interacting 
potential defined as
\begin{equation}
\label{eq5}
V_{int} = \big[V_0\hat{I}+\boldsymbol{\sigma}.\mathbf{V}_m + 
V_R \hat{e}_x.(\hat{\boldsymbol{\sigma}}\times\mathbf{k})\big]\big[\delta(x)+\delta(x-L)\big]
\end{equation}
\noindent where, $V_0$ gives the intrinsic spin independent barrier potential while $\mathbf{V}_m$ represent the spin dependent the barrier potential. We 
define the spin dependent potential as \cite{linder12} 
\begin{align}\notag
\mathbf{V}_m & =(V_x,V_y,V_z)\\\notag
 & =(\rho V_0 \cos\xi_m\sin\chi_m, 
\rho V_0 \sin\xi_m\sin\chi_m, \rho V_0 \cos\chi_m)
\end{align} 
where, $\rho = |\mathbf{V}_m|/V_0$ represent the effective ratio between the 
barrier magnetic and non-magnetic moments. The last term in the 
first square backet of Eq.~\eqref{eq5} is the contribution due to the RSOC, 
with $V_R$ being the strength of the RSOC and it arises due to the asymmetry 
of the interface as well as due to NCSC \cite{acharjee2,acharjee3,badica,
linder15,frigeri}.

Since both even and odd frequency correlations are present in NCSC at the same 
time, so the gap matrix $\hat{\Delta }_{\alpha \beta}$ appearing in 
Eq.~(\ref{eq2}) can be represented as a superposition of the singlet and 
triplet components \cite{acharjee2,acharjee3,linder15}, i.e.
\begin{multline}
\label{eq6}
 \hat{\Delta }_{\alpha \beta } = 
\left(
\begin{array}{cc}
 \Delta^T _{k\uparrow\uparrow} & \Delta^S _{k\uparrow\downarrow}+\Delta^T _{k\uparrow\downarrow} \\
-\Delta^S _{k\uparrow\downarrow}+\Delta^T _{k\uparrow\downarrow} & \Delta^T _{k\downarrow\downarrow} \\
\end{array}
\right)\\
\times\big[\Theta(-x)e^{i\phi_L}+\Theta(x-L)e^{i\phi_R}\big]
\end{multline}
where, $\Delta^T _{k\sigma\sigma'}$ and $\Delta^S _{k\sigma\sigma'}$ are 
contributions due to the singlet (S) and triplet (T) components respectively. 
Here, $\phi_{L}$($\phi_{R}$) represents the macroscopic phase of left(right) 
superconductor. Considering the RSOC vector for CePt$_3$Si \cite{linder15,
frigeri,acharjee2}  $\mathbf{g_k} = V_R(k_y,-k_x,0)^T$, the gap vectors can 
be defined as 
\begin{multline}\nonumber 
\Delta^T_{k\uparrow\uparrow}(\Delta^T_{k\downarrow\downarrow}) = \mp \frac{\Delta_t}{2|\mathbf{k}|}(k_y \pm ik_x)\big[\Theta(-x)e^{\mp i\phi_L}\\
+\Theta(x-L)e^{\mp i\phi_R}\big]
\end{multline} 
and $\Delta^S _{k\uparrow\downarrow}=\Delta_k$. We define a parameter 
$\delta$ to characterize singlet-triplet mixing as $\Delta_s=\Delta_0\delta$ 
and $\Delta_t=\Delta_0(1-\delta)$. So, in view of Eqs.~(\ref{eq3}), (\ref{eq4}),
(\ref{eq5}) and (\ref{eq6}), the Hamiltonian \eqref{eq2} can be written as
 \begin{equation}
\label{eq7}
\hat{\mathcal{H}} = \left(
\begin{array}{cccc}
 h_z+\mathcal{H'} & g_{k_-}+\Gamma_{xy} & \Delta_{k\uparrow\uparrow} & \Delta_k \\
 g_{k_+}+\Gamma^*_{xy} &  -h_z+\mathcal{H'} & -\Delta_k & \Delta _{k\downarrow\downarrow} \\
 {\Delta_{k\uparrow\uparrow}}^\dagger &  -{\Delta_{k}}^\dagger & -h_z-\mathcal{H}' & g_{k_+}-\Gamma^*_{xy} \\
 {\Delta_{k}}^\dagger &  {\Delta_{k\downarrow\downarrow}}^\dagger & g_{k_-}-\Gamma_{xy} & h_z-\mathcal{H}' \\
\end{array}
\right)
  \end{equation}
\noindent where, $\mathcal{H}' = -\mu_i[\Theta(-x) + \Theta(x-L)] + (V_0+\sigma V_z)[\delta(x)+\delta(x-L)]$, $\Gamma_{xy} = v_F(k_x-ik_y) + h_{xy}$, $h_{xy}=(m_x-im_y)[\Theta(x)\Theta(L-x)]+(V_x-iV_y)[\delta(x)+\delta(x-L)]$, $h_z = m_z[\Theta(x)\Theta(L-x)]$ and 
$g_{k\pm} = V_R(k_x \mp i k_y)[\Theta(-x)e^{\mp i\phi_L}+\Theta(x-L)e^{\mp i\phi_R}]$.\\

Assuming $\theta$ as the incident angle of the quasi particle in the left NCSC, 
the trajectory of the particle can be expressed using the momentum components, 
$k_x = |\mathbf{k}|\cos\theta$ and $k_y = |\mathbf{k}|\sin\theta$. Thus
the total wave function of an electron injected in the left NCSC with an angle 
$\theta$ can be written as
\begin{multline}
\label{eq8}
\Psi_{\text{T}}(x) = \big[\Psi^{\text{L}}_{\text{NCSC}}(x)\Theta(-x)+\Psi_{\text{HM}}(x)\Theta(x)\Theta(L-x)
\\+\Psi^{\text{R}}_{\text{NCSC}}(x)\Theta(x-L)\big]e^{ik_yy}
\end{multline}
where, $k_y$ is the momentum parallel to the interface and 
$\Psi^{\text{L}}_{\text{NCSC}}(x)$, $\Psi_{\text{HM}}(x)$, 
$\Psi^{\text{R}}_{\text{NCSC}}(x)$ are the quasi particle wave functions in 
the left NCSC, half metal and right NCSC respectively. The wave functions in 
different layers are obtained by diagonalizing the Hamiltonian (\ref{eq7}) and 
can be written as \cite{acharjee3}
\begin{multline}
\label{eq9}
\Psi^{\text{L}}_{\text{NCSC}}(x<0) = \\
\tau[u_\pm\hat{\eta}_1 \pm u_\pm e^{-i\phi_L}\hat{\eta}_2 
\mp v_\pm e^{-i\phi_L}\hat{\eta}_3 + v_\pm\hat{\eta}_4] e^{ik^+_{x\pm}x}\\
a_1[u_+\hat{\eta}_1 + u_+ e^{-i\phi_L}\hat{\eta}_2 
- v_+ e^{-i\phi_L}\hat{\eta}_3 + v_+\hat{\eta}_4] e^{-ik^+_{x+}x}
\\
+ a_2[u_-\hat{\eta}_1 - u_- e^{-i\phi_L}\hat{\eta}_2
+ v_- e^{-i\phi_L}\hat{\eta}_3 + v_-\hat{\eta}_4] e^{-ik^+_{x-}x}
\\
+ a_3[v_+\hat{\eta}_1 + v_+ e^{-i\phi_L}\hat{\eta}_2
- u_+ e^{-i\phi_L}\hat{\eta}_3 + u_+\hat{\eta}_4] e^{ik^-_{x+}x}
\\
+ a_4[v_-\hat{\eta}_1 - v_- e^{-i\phi_L}\hat{\eta}_2
+ u_- e^{-i\phi_L}\hat{\eta}_3 + u_-\hat{\eta}_4] e^{ik^-_{x-}x}.
\end{multline}
The wave function in the right NCSC region in a similar way can be written as
\begin{multline}
\label{eq10}
\Psi^{\text{R}}_{\text{NCSC}}(x > L) = \\
c_1[u_+\hat{\eta}_1 + u_+ e^{-i\phi_R}\hat{\eta}_2 
- v_+ e^{-i\phi_R}\hat{\eta}_3 + v_+\hat{\eta}_4] e^{ik^+_{x+}x}
\\
+ c_2[u_-\hat{\eta}_1 - u_- e^{-i\phi_R}\hat{\eta}_2
+ v_- e^{-i\phi_R}\hat{\eta}_3 + v_-\hat{\eta}_4] e^{ik^+_{x-}x}
\\
+ c_3[v_+\hat{\eta}_1 + v_+ e^{-i\phi_R}\hat{\eta}_2
- u_+ e^{-i\phi_R}\hat{\eta}_3 + u_+\hat{\eta}_4] e^{-ik^-_{x+}x}
\\
+ c_4[v_-\hat{\eta}_1 - v_- e^{-i\phi_R}\hat{\eta}_2
+ u_- e^{-i\phi_R}\hat{\eta}_3 + u_-\hat{\eta}_4] e^{-ik^-_{x-}x}.
\end{multline}
Here we define, $\hat{\eta}_1 = (1,0,0,0)^T$, $\hat{\eta}_2 = (0,1,0,0)^T$, 
$\hat{\eta}_3 = (0,0,1,0)^T$ and
 $\hat{\eta}_4 = (0,0,0,1)^T$. Here,  $a_1$($a_2$) are the reflection 
coefficients for up(down) spin electrons, while $a_3$($a_4$) are the 
reflection coefficients for the up(down) spin holes. $c_1$($c_2$) are the 
transmission coefficients for up(down) spin electrons, while   $c_3$($c_4$) 
are the transmission coefficients for the up(down)spin holes in right NCSC 
region. $\phi_L$ and $\phi_R$ are the superconducting phase factor for 
the left and right NCSC respectively. $k^{+(-)}_{x\sigma}$ gives the momenta 
of the electron (hole) in NCSC with $\sigma = \pm 1$ represent spin up and 
down particles. The momenta of the electron and the hole are defined as 
 \begin{equation}
\label{eq11}
k^\pm_{x\sigma} = \sqrt{2(\mu_S\pm\Omega_{\pm}) - \sigma Z_R}
\end{equation}
where, $\mu_S$ is the chemical potential and $Z_R$ is strength of RSOC in NCSC 
region. The factor $\Omega_{\pm}$ is defined as 
\begin{equation}
\label{eq12}
\Omega_{\pm} = \sqrt{E^2-|\Delta_s\pm\frac{\Delta_t}{2}|^2}.
\end{equation}
The quasi particle amplitudes $u_\pm$ and $v_\pm$ appearing in Eqs.~(\ref{eq9}) 
 and (\ref{eq10}) are defined as
\begin{eqnarray}
\label{eq13}
\label{eq14}
 u_\pm =  \frac{1}{\sqrt{2}}\sqrt{1+\frac{\Omega_{\pm}}{E}}\\
 v_\pm =  \frac{1}{\sqrt{2}}\sqrt{1-\frac{\Omega_{\pm}}{E}}
\end{eqnarray}

In a similar way the wave function in the half metallic region is \cite{linder12}
\begin{multline}
\label{eq15}
 \Psi_{\text{HM}}(0 < x < L) = b_1\hat{\eta}_1e^{i\kappa^+_{x+}x} + b_2\hat{\eta}_2e^{i\kappa^+_{x-}x} 
\\ + b_3\hat{\eta}_1e^{-i\kappa^-_{x+}x} + b_4\hat{\eta}_2e^{-i\kappa^-_{x-}x}
\end{multline}
where, $b_1$, $b_2$, $b_3$, $b_4$ are the transmission coefficients for 
electron and hole in half metallic region. The quasi particle momenta in this 
region is given by
\begin{multline}
\label{eq16}
\kappa^\pm_{x\sigma} =\pm\mu_{TI}\mp\sqrt{M_z^2+(h_x+k_xv_F)^2 + (h_y+k_yv_F)^2}
\end{multline}
where, $M_z = h_z+V_z$ and $\mu_{TI}$ is the chemical potential of TI. In most 
of the TIs, chemical potential lies closer to the valence band as they are not 
perfectly insulating. So in experiment \cite{beenakker}, the chemical potential
has larger value than the superconducting gap. The band dispersion at the 
surface of topological insulator is plotted using Eq.~(\ref{eq16}) and 
displayed in Fig.~(\ref{fig2}). It is noted that there exist a gap in the band 
energy when both bulk and barrier magnetic moment are parallel to each other 
along z-direction with $\rho = 0.3$. However, the energy gap disappears for 
the anti-parallel orientation. This signifies that the band dispersion at the 
surface of 3D TIs depends on the alignment angle of the magnetic moment due to 
bulk HM and barrier. The boundary conditions that are to be satisfied by 
the wavefunctions at two junctions are as follows:
\begin{align}
\label{eq17}
\big[\Psi^{\text{L}}_\text{NCSC}(x) - \Psi_\text{HM}(x)\big]\big|_{x\,=\,0}& = 0,\\
\label{eq18}
\big[\Psi_\text{HM}(x)-\Psi^{\text{R}}_\text{NCSC}(x)\big]\big|_{x\,=\,L}& = 0,\\
\label{eq19}
\partial_x\big[\Psi_\text{HM}(x)-\Psi^{\text{L}}_\text{NCSC}(x)\big]|_{x\,=\,0} 
& = \big(\hat{\tau}+\hat{\Lambda}_m\big)\Psi^{\text{L}}_\text{NCSC}(x)\big|_{x\,=\,0},\\
\label{eq20}
\partial_x\big[\Psi^{\text{R}}_\text{NCSC}(x)-\Psi_\text{HM}(x)\big]|_{x\,=\,L} &= \big(\hat{\tau}+\hat{\Lambda}_m\big)\Psi_\text{HM}(x)\big|_{x\,=\,L}
\end{align}
where, 
$\hat{\tau}$ is defined as \cite{klam}
 \begin{equation}
\label{eq21}
\hat{\tau} = \left(
\begin{array}{cccc}
Z_- & 0 & 0 & 0 \\
 0 &  Z_+ & 0 & 0 \\
0 &  0 & Z_+ & 0 \\
 0 &  0 & 0 & Z_- \\
\end{array}
\right)
\end{equation}
with $Z_\pm = Z_0 \pm Z_R \sin\theta$ where, $Z_0=\frac{2V_0}{k_F\cos\theta}$ 
and $Z_R=2V_R\tan\theta$ are the parameters characterizing spin independent 
barrier strength and RSOC. The matrix $\hat{\Lambda}_m$ gives the contribution 
due to spin active interacting potential and has the following 
form \cite{linder16}:
 \begin{equation}
\label{eq22}
\hat{\Lambda}_m = \left(
\begin{array}{cccc}
\lambda_1 & \lambda_2 & 0 & 0 \\
 \lambda_2^* &  -\lambda_1 & 0 & 0 \\
0 &  0 & \lambda_1& \lambda_2^* \\
 0 &  0 & \lambda_2 & -\lambda_1 \\
\end{array}
\right)
\end{equation}
where, $\lambda_1 = 2\rho V_0\cos\chi_m$ and $\lambda_2 = 2\rho V_0\sin\chi_m e^{-i\xi_m}$.

\section{Andreev Bound States and Josephson Supercurrent}
The wave functions must satisfy the boundary conditions defined in 
Eqs.~(\ref{eq17}), (\ref{eq18}), (\ref{eq19}) and (\ref{eq20}). This leads to 
a system of $16$ linear equations connecting the coefficients defined as 
$\hat{A}\hat{x} = 0$, where $\hat{A}$ is a $16\times16$ matrix and 
$\hat{x} = (a_1,a_2,a_3,a_4,b_1,b_2,b_3,b_4,b_1',b_2',b_3',b_4',c_1,c_2,c_3,c_4)^T$. The Andreev Bound States (ABS) can be calculated by the condition 
$det(\hat{A}) = 0$. The solution for these equations needed several lengthy 
calculations. However, under some limiting cases they are expressible in 
analytic form. Considering the limiting cases, like normal interface 
($\theta = 0$), transparent barrier ($Z_0 = 0$), Rashba free case ($Z_R = 0$) 
and small length scale ($L/L_0 << 1$), the Andreev levels can be expressed 
as
\begin{equation}
\label{eq23}
E_{\sigma\tau} = \tau\beta\cos\left(\frac{\phi+\sigma\chi_m}{2}\right)
\end{equation} 
where, $\tau = \pm 1$ for left and right moving Cooper pairs respectively, 
$\sigma$ is the helicity index and $\beta = \sqrt{\Gamma_1^2+\Gamma_2^2+\Gamma_3^2}$. The parameters $\Gamma_1$, $\Gamma_2$ and $\Gamma_3$ are defined as  
\begin{multline}
\Gamma_1 = \big[\rho\lbrace\cos\chi_m(\rho\cos\chi-2)-\rho\sin^2\xi_m\sin^2\chi_m\rbrace
\\+2\big]\Delta_-^2-\sqrt{2}\Delta_+^2\sqrt{\Gamma_4\rho^2\sin^2\xi_m\sin^2\chi_m},
\end{multline}
\begin{multline}
\Gamma_2 = 2\Delta_-^2\rho\sin\xi_m\sin\chi_m-\sqrt{2}\Delta_+^2\times\\
\sqrt{\Gamma_4\rho^3\sin^3\xi_m\sin ^4\chi_m},
\end{multline}
\begin{equation}
\Gamma_3 = \frac{2\Gamma_4\rho^2\sin^2(\xi_m\sin ^2\chi_m}{2\rho\cos\chi_m(\rho\cos\chi_m+1)+1}
\end{equation}
with
\begin{equation}
\Gamma_4 = \rho^2\cos2\chi_m+\rho^2-4\rho\cos\chi_m+4.
\end{equation}

Using the approximated analytical expression in Eq.~(\ref{eq23}), we study the 
formation of ABS and its dependence on the parameters like barrier 
spin-dependent parameters ($\rho$, $\chi_m$, $\xi_m$) and singlet-triplet 
mixing parameter 
($\delta$). However, if we consider the barrier thickness ($Z_0$), RSOC 
strength ($Z_R$) and oblique incident angle ($\theta$), then the calculations 
become a too cumbersome and the analytical expression for the same is too hard 
to obtain. So numerical approach is followed to obtain ABS and the results are 
also displayed in the later sections.

The angle-averaged Josephson supercurrents of the ABS are evaluated using the 
standard expression \cite{linder12},
\begin{equation}
\label{eq24}
\frac{J}{J_0} = \int_{-\pi/2}^{\pi/2}d\theta\cos\theta\tanh\left(\frac{E_{\sigma\tau}}{2k_BT}\right)\frac{dE_{\sigma}/\Delta_{\sigma\tau}}{d\phi}
\end{equation}  
where, $J_0 = \frac{eN\Delta_\sigma}{\hbar}$. We obtain the numerical plots of 
Josephson supercurrents by considering $T/T_c = 0.001$ with $T_c$ being the 
critical temperature. We set, $v_F=1$, $h_{z}=1$, $h_{xy} = 0$ and as  
$\Delta_0 << \mu_{S,TI}$ \cite{beenakker}, so we consider 
$\mu_S = 150\Delta_0$ and $\mu_{TI} = 100\Delta_0$ for all our analysis. 
 
\begin{figure*}[hbt]
\centerline
\centerline{
\includegraphics[scale = 0.47]{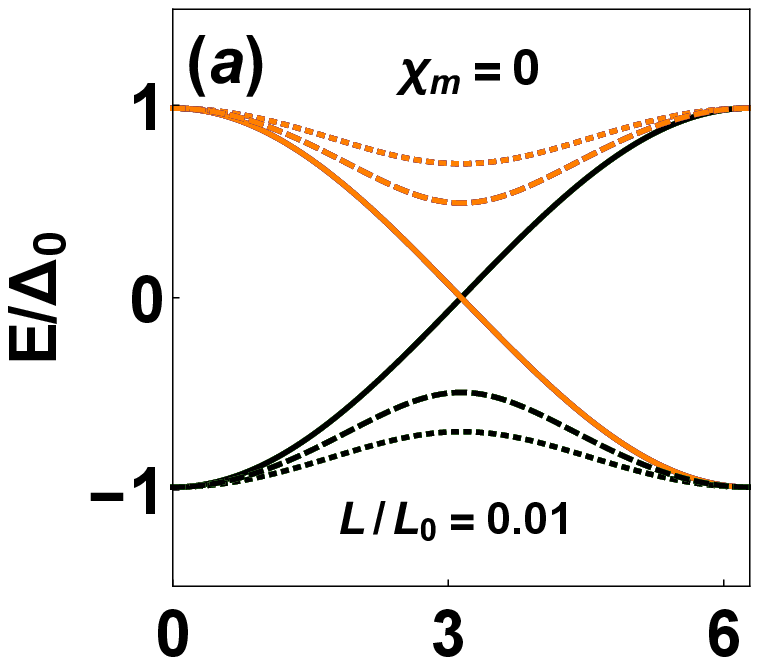}
\hspace{0.001mm}
\vspace{0.1cm}
\includegraphics[scale = 0.47]{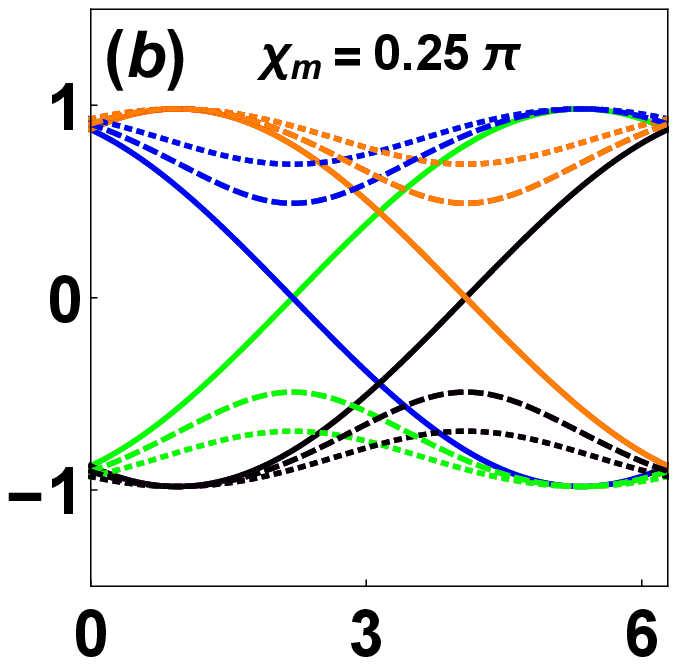}
\hspace{0.001mm}
\vspace{0.1cm}
\includegraphics[scale = 0.47]{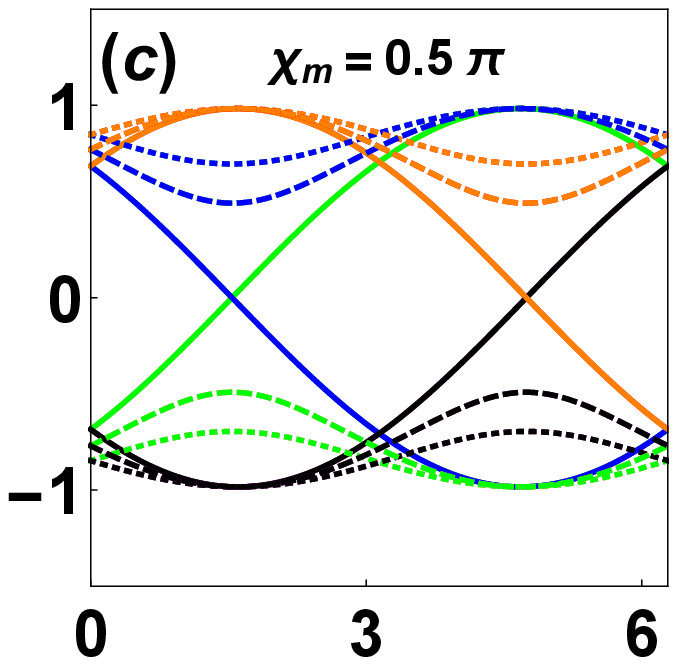}
\hspace{0.001mm}
\vspace{0.1cm}
\includegraphics[scale = 0.47]{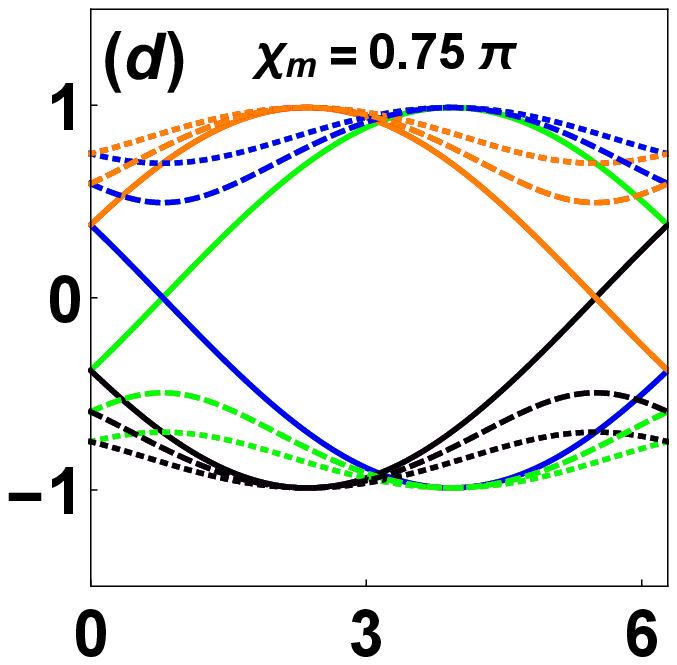}
}
\hspace{0.001mm}
\vspace{0.1cm}
\includegraphics[scale = 0.47]{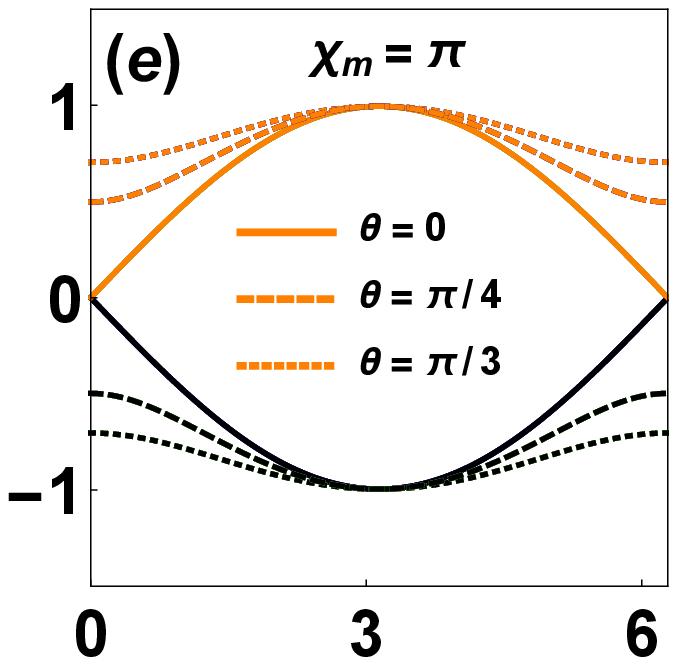}
\hspace{0.001mm}

\includegraphics[scale = 0.47]{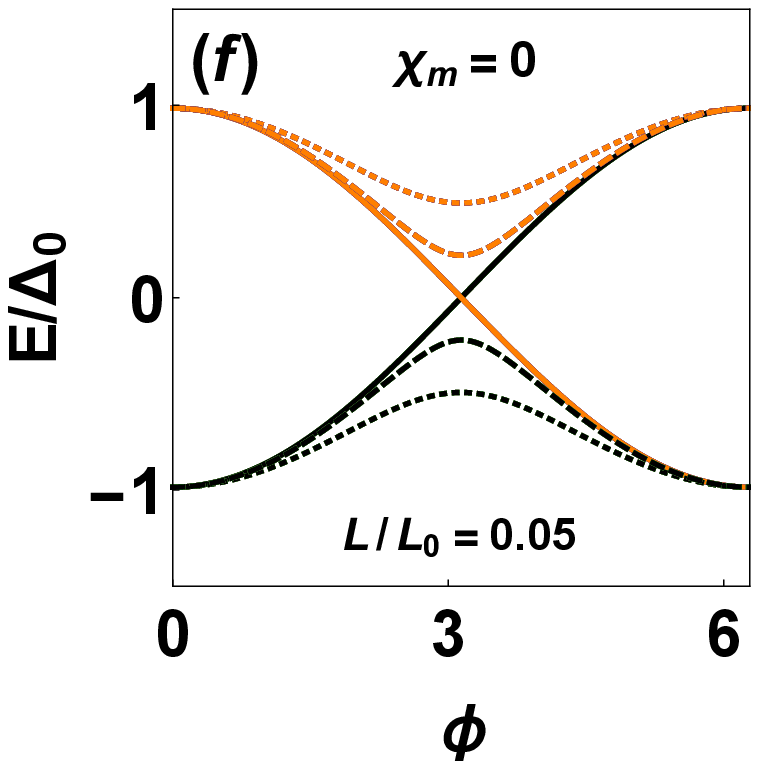}
\hspace{0.001mm}
\includegraphics[scale = 0.47]{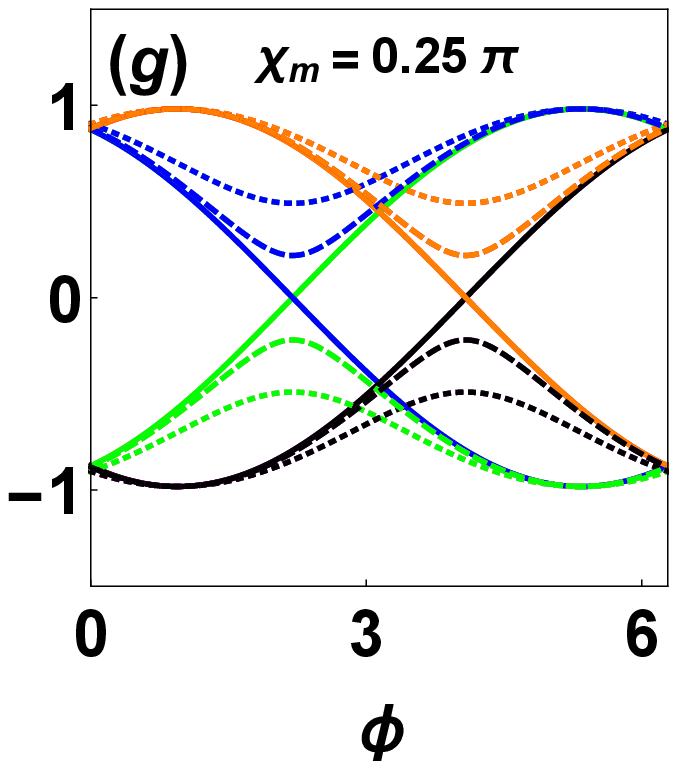}
\hspace{0.001mm}
\includegraphics[scale = 0.47]{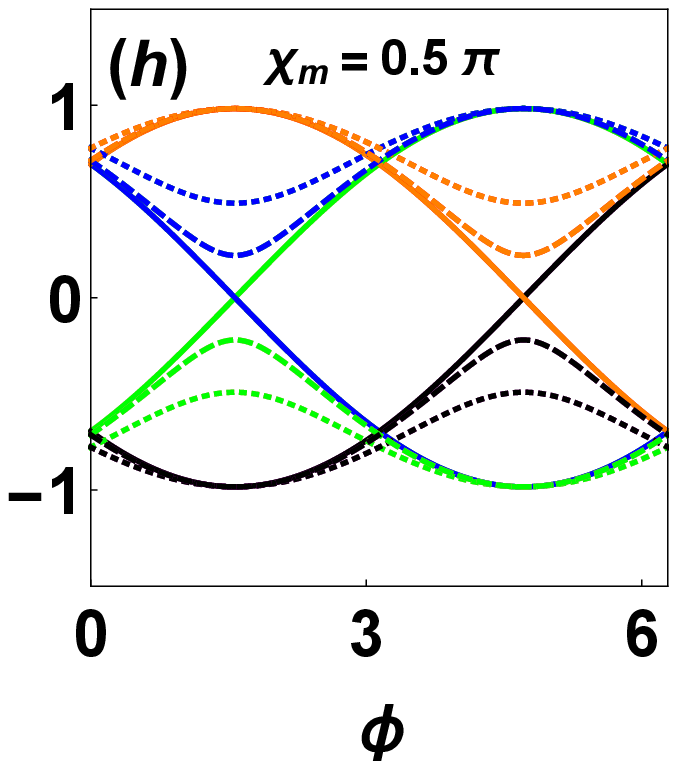}
\hspace{0.001mm}
\includegraphics[scale = 0.47]{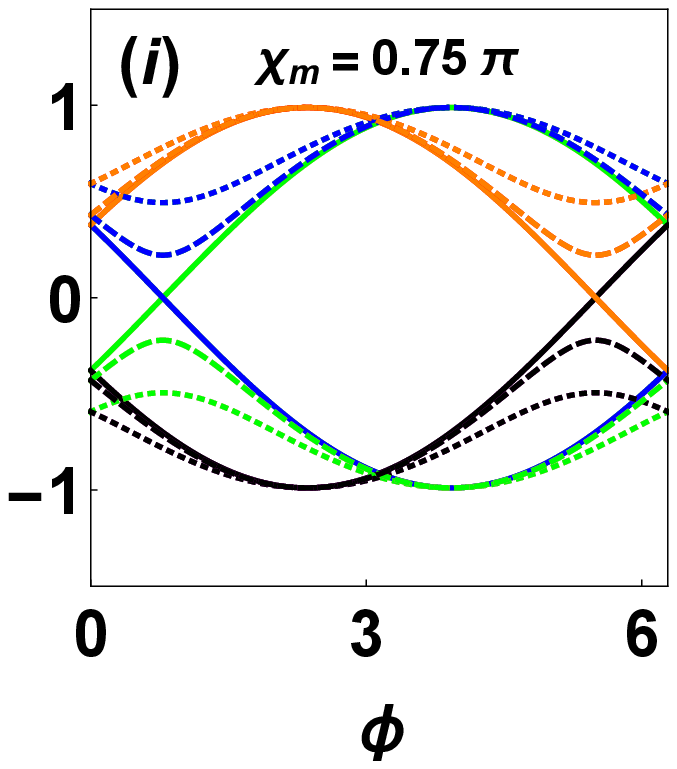}
\hspace{0.001mm}
\includegraphics[scale = 0.47]{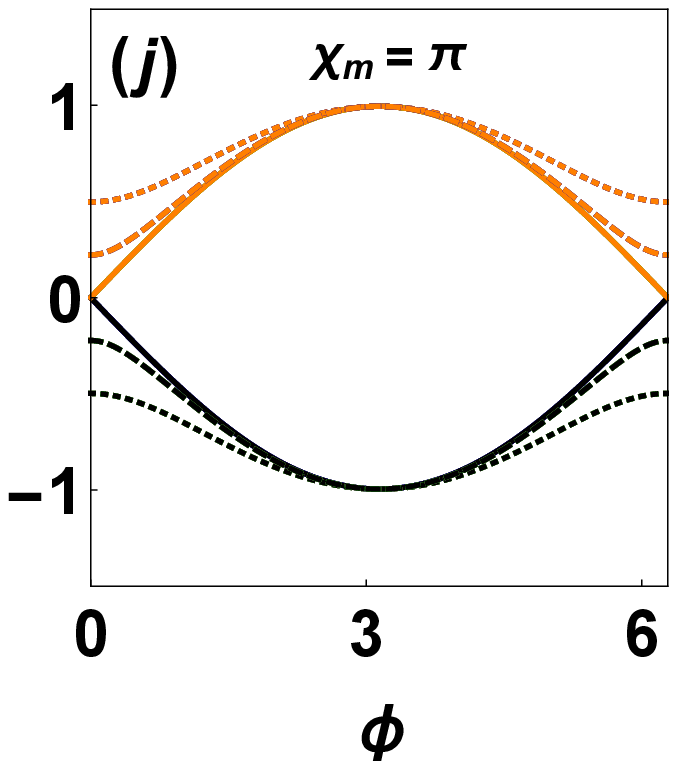}
\caption{ABS spectra for different choices of angle of incidence 
($\theta$) and polar angle ($\chi_m$) considering $\rho=0.01$, $\xi_m=0.5\pi$, 
$\delta = 1$, $Z_R = 0$ and $Z_0 = 0.1$. The plots in the top panel are for 
$L/L_0 = 0.01$ while the plots in the bottom panel are for $L/L_0 = 0.05$.}
\label{fig3}
\end{figure*} 
\section{Results and Discussions}
\subsection{Andreev Bound States}
First, we consider the ABS corresponding to the different angles of incidence 
of the injected particles. The results are displayed in Fig.~\ref{fig3} for 
different misalignment angles considering two different length scales viz., 
$L/L_0 = 0.01$ and $0.05$. We consider $Z_0 = 0.1$, $Z_R = 0$ and 
$\delta = 1$ for this analysis. We find a gapless dispersion with $4\pi$ 
periodic ABS for perpendicular incidence as represented by the solid curves of 
Fig.~\ref{fig3}(a). It indicates the presence of Majorana at $\phi = 
\pi$ for perpendicular incidence. More interestingly result in 
Fig.~\ref{fig3}(a) signifies that Majorana is present in a spin-active barrier.
However, for oblique incidence, a gap is always observed for all cases, as 
shown by the dotted and dashed lines of Figs.~\ref{fig3}(a)-\ref{fig3}(j). It 
is to be noted that the gap opens up further for larger the angle of incidence 
as seen for $\theta = \frac{\pi}{3}$ indicated by the dotted lines. The 
presence of a gap at oblique incidence is due to non-zero backscattering at a 
finite angle. To realize the impact of barrier magnetic moment, we have 
plotted ABS in Figs.~\ref{fig3}(a)-\ref{fig3}(j) for different values of 
$\chi_m$ with $\rho=0.01$ and $\xi_m=0.5\pi$. It is seen that the Andreev 
levels split corresponding to spin helicity $\sigma$ and for non-vanishing 
values of $\chi_m$. The wave vector mismatch occurs for the misalignment of 
the polar angle of magnetic moments. Thus the two degenerate energy bands 
split into two more branches, as seen in Figs.~\ref{fig3}(b)-\ref{fig3}(d) 
and Figs.~\ref{fig3}(g)-\ref{fig3}(i). We also find that the two sets of ABS 
depart away from each other with the increase in misalignment angle $\chi_m$. 
This results in the formation of a $\phi$ junction for non-vanishing values of 
$\chi_m$ corresponding to normal incidence. Similar behavior is also observed 
for oblique incidence. However, the gap remains constant in this situation 
also. It is seen that the $4\pi$ periodicity of the ABS will not present for 
non-vanishing values of $\chi_m$ and finally for anti-parallel orientation 
$\chi_m=\pi$, the system resides in $2\pi$-ABS branches as seen from 
Figs.~\ref{fig3}(e) and \ref{fig3}(j). Thus, interaction with the barrier 
magnetic moments results in the system to reside in the lower Andreev energy 
branches. The Andreev levels are found to display similar characteristics in 
Figs.~\ref{fig3}(f)-\ref{fig3}(j) for $L/L_0 = 0.05$ in normal incidence 
condition. However, for oblique incidence, the gap is significantly reduced 
for this length scale. It may be due to the reason that with increasing 
length, induced magnetic proximity on the surface of 3D TI gets weaker and 
hence the effect of barrier and bulk moment is reduced. Thus, the system is 
found to display $4\pi$ periodicity for the parallel alignment while it shows 
anomalous characteristics and $\phi$ periodicity for the misalignment of the 
barrier and bulk moments.  
      
\begin{figure*}[hbt]
\centerline
\centerline{
\includegraphics[scale = 0.47]{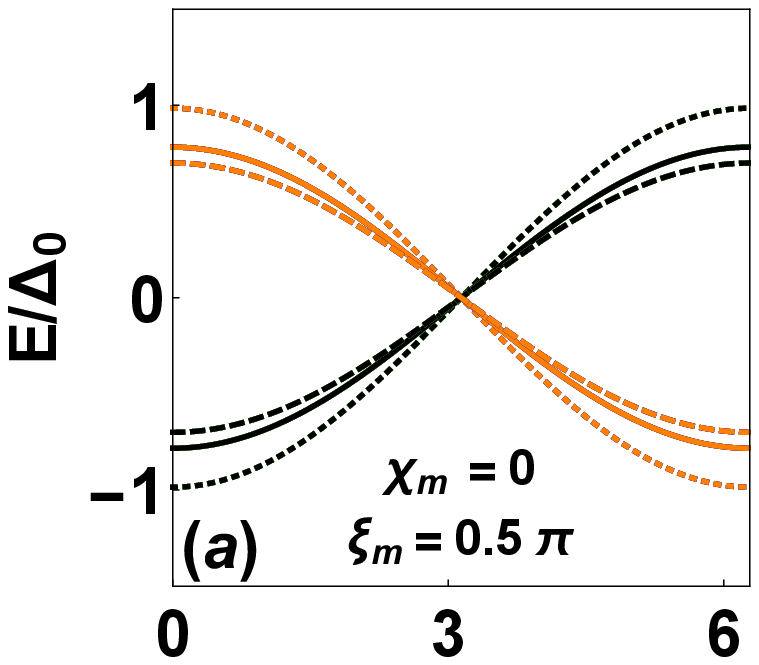}
\hspace{0.001mm}
\vspace{0.1cm}
\includegraphics[scale = 0.47]{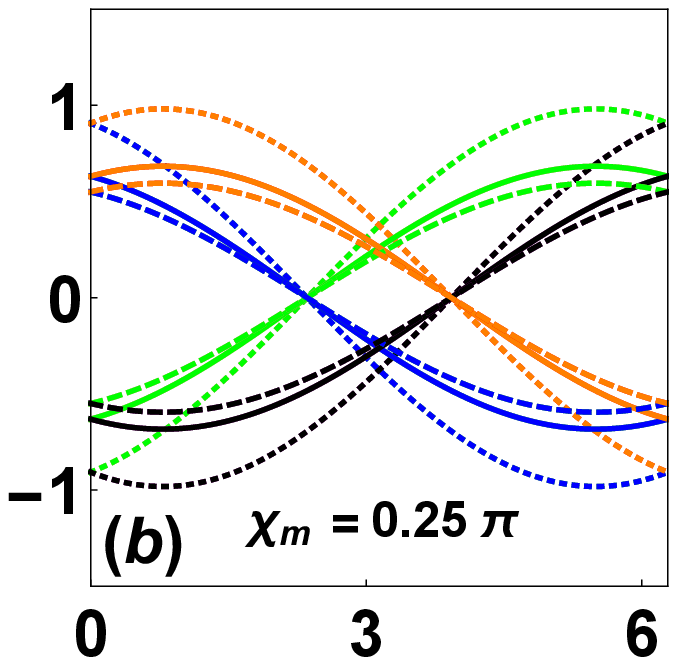}
\hspace{0.001mm}
\vspace{0.1cm}
\includegraphics[scale = 0.47]{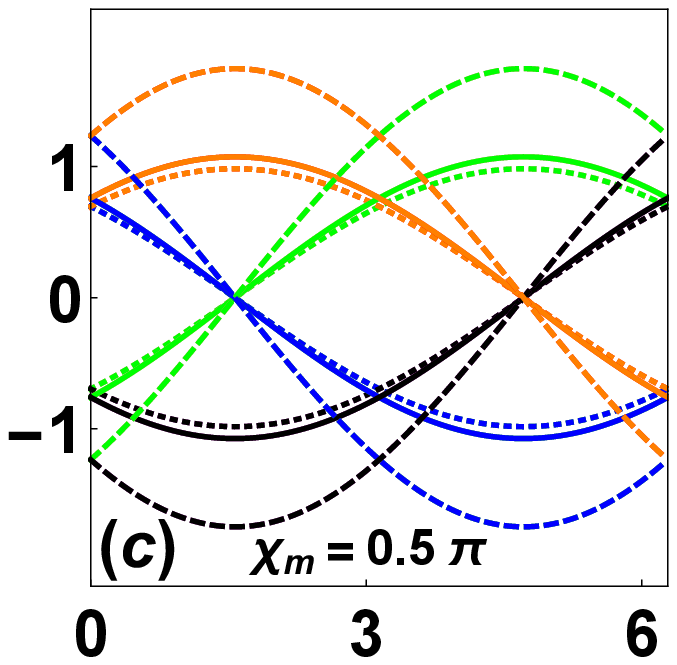}
\hspace{0.001mm}
\vspace{0.1cm}
\includegraphics[scale = 0.47]{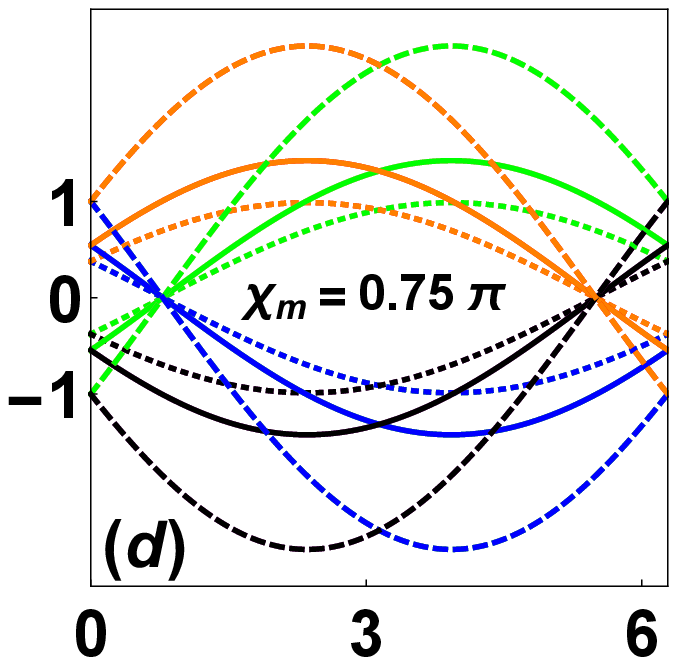}
}
\hspace{0.001mm}
\vspace{0.1cm}
\includegraphics[scale = 0.47]{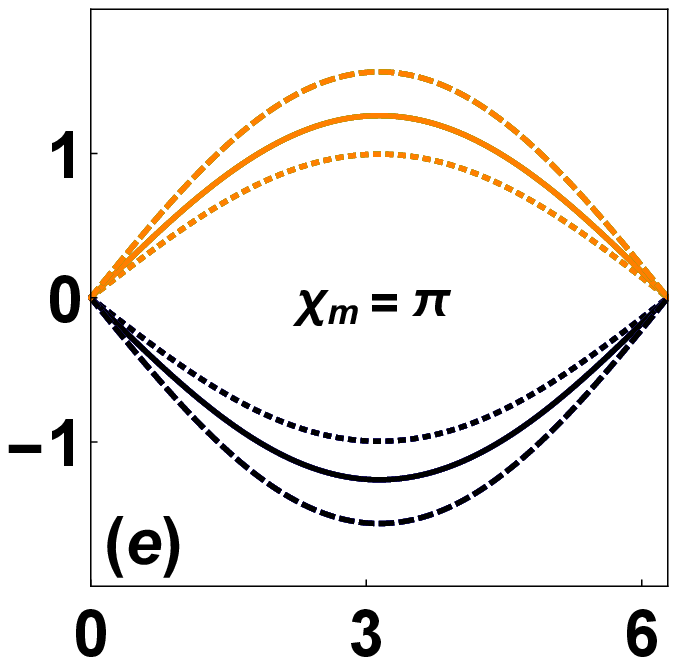}
\hspace{0.001mm}

\includegraphics[scale = 0.47]{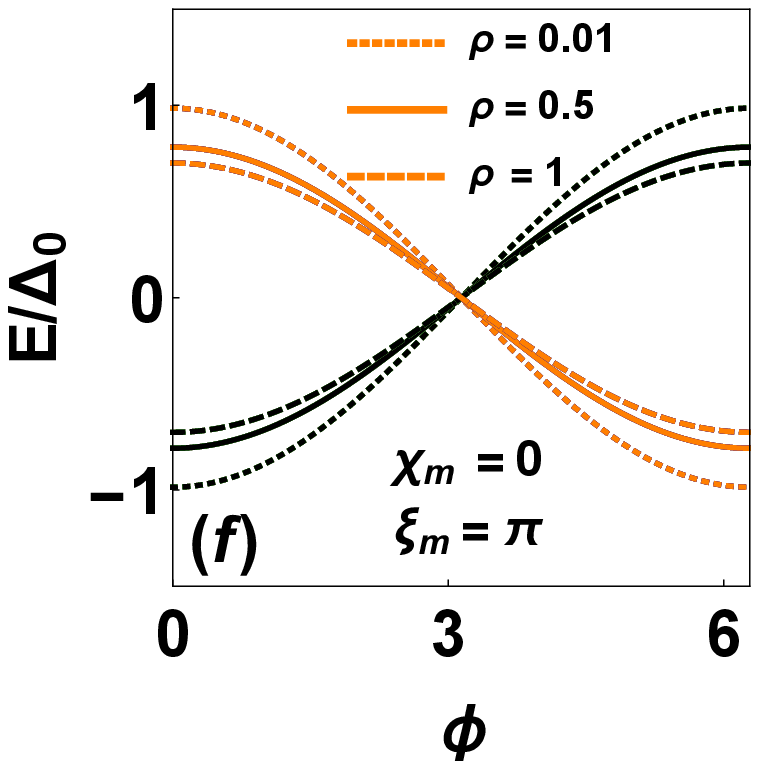}
\hspace{0.001mm}
\includegraphics[scale = 0.47]{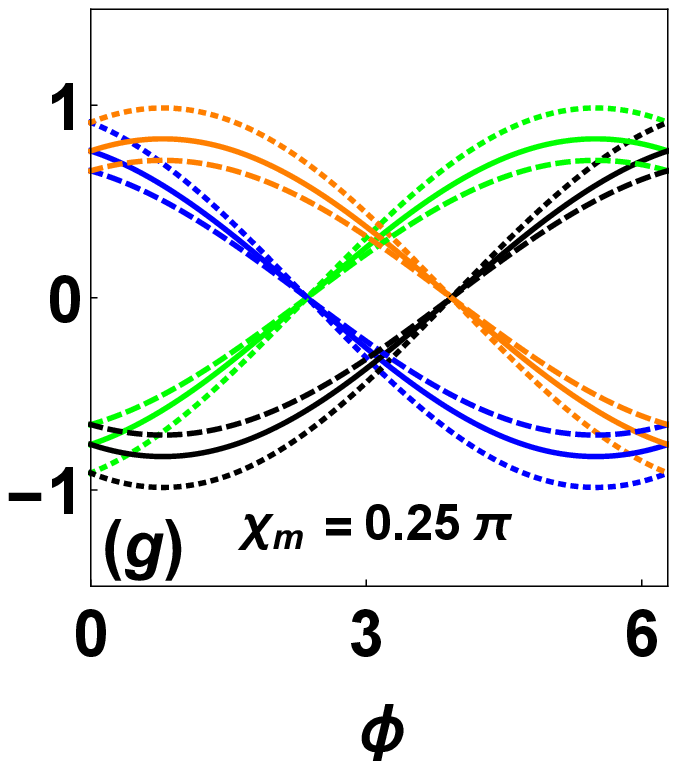}
\hspace{0.001mm}
\includegraphics[scale = 0.47]{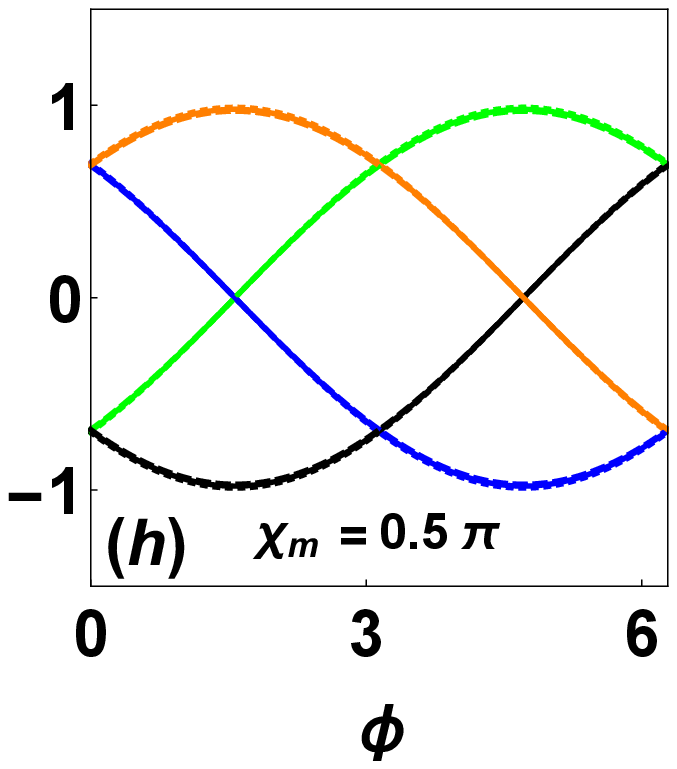}
\hspace{0.001mm}
\includegraphics[scale = 0.47]{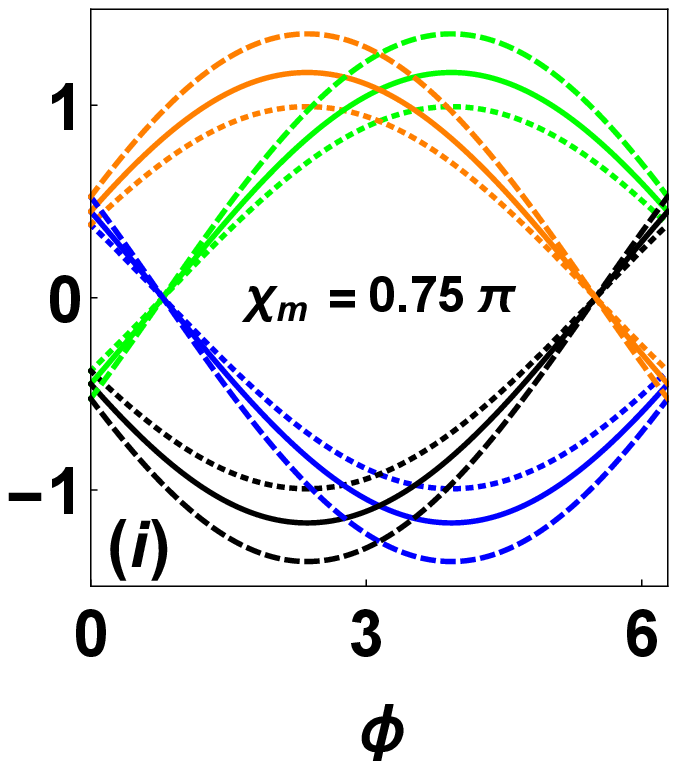}
\hspace{0.001mm}
\includegraphics[scale = 0.47]{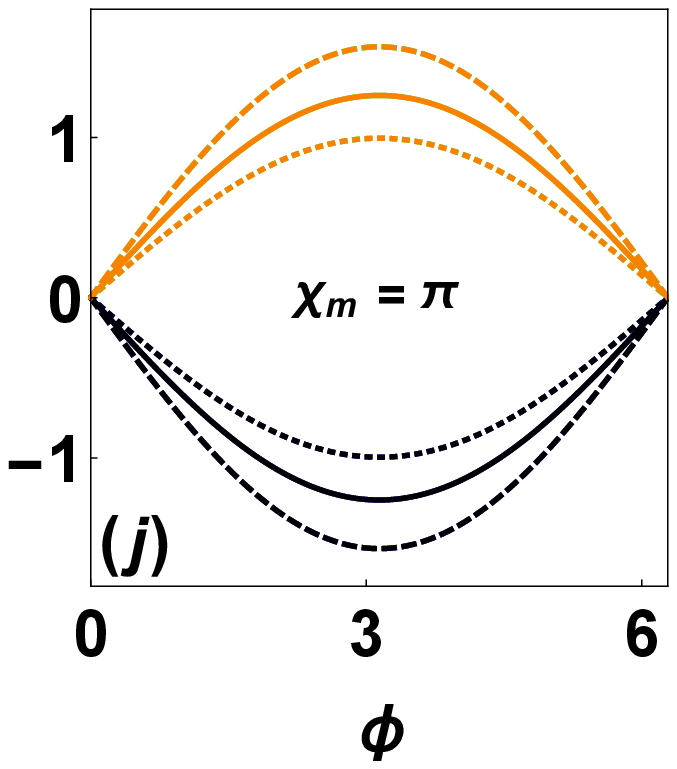}
\caption{ABS spectra for different choices of $\rho$ and polar angle 
($\chi_m$) considering $\delta = 1$, $\theta = 0$, $Z_R = 0$ and $Z_0 = 0.1$. 
The plots in the top panel are for $\xi_m=0.5\pi$ while the plots in the 
bottom panel are for $\xi_m=\pi$.}
\label{fig4}
\end{figure*}

In Figs.~\ref{fig4}(a)-\ref{fig4}(j), we show the dependence of ABS on the 
ratio of barrier magnetic to non-magnetic moment ($\rho$) in perpendicular 
incidence condition. We study the ABS spectra for different polar angles 
($\chi_m$). The plots in the top panel are for azimuthal angle $\xi_m = 0.5\pi$
while the plots in bottom panel are for $\xi_m = \pi$. We 
find that the Majorana modes appear at $\phi = \pi$  in parallel configuration 
for all choices of $\rho$ and $\xi_m$ as seen in Figs.~\ref{fig4}(a) and 
\ref{fig4}(f). Andreev levels shrinks with the increasing values of $\rho$ and 
maximum contraction is observed for $\rho=1$ as seen from the dashed curves of 
Figs.~\ref{fig4}(a) and \ref{fig4}(b). However, for $\chi_m = 0.5\pi$, 
$0.75\pi$ and $\pi$ the ABS grows more for $\rho=1$ and a opposite 
characteristic is seen. It signifies that for anti-parallel alignment ABS 
energy becomes maximum when magnitude of barrier magnetic moment approaches to 
non-magnetic moment. This behaviour is seen in 
Figs.~\ref{fig4}(c)-\ref{fig4}(e) for $\xi_m = 0.5\pi$. The ABS depart again 
for non-vanishing values of $\chi_m$. However, the Andreev levels for 
$\rho=0.01$ and $0.5$ are found to display nearly same characteristics for 
$\chi_m = 0.5\pi$. The ABS shrinks further for $\chi_m = 0.75\pi$ with 
$\rho=0.01$ as seen from Fig.~\ref{fig4}(d). Though a similar characteristic 
is seen for $\chi_m = 0$ and $0.25\pi$ with $\xi_m=\pi$, but for 
$\chi_m = 0.5\pi$ the Andreev energy levels are found to be independent of 
$\rho$ as seen from Fig.~\ref{fig4}(h). For $\chi_m = 0.75\pi$ and $\pi$  with 
$\xi_m = \pi$, the Andreev levels grows for all choices of $\rho$. Thus, 
although the Andreev energy levels becomes anomalous for nonzero values of 
$\chi_m$ but the magnitude of the Andreev levels can significantly change 
for different azimuthal angle as seen from Fig.~\ref{fig4}. This indicate that 
the ABS not only depends on the polar angle $\chi_m$ but also on azimuthal 
angle $\xi_m$.

\begin{figure*}[hbt]
\centerline
\centerline{
\includegraphics[scale = 0.47]{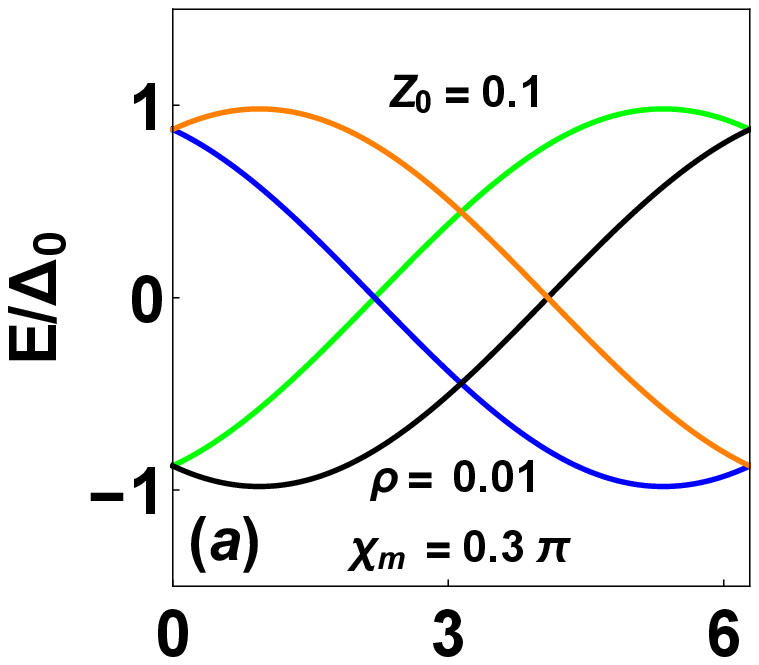}
\hspace{0.01cm}
\vspace{0.1cm}
\includegraphics[scale = 0.47]{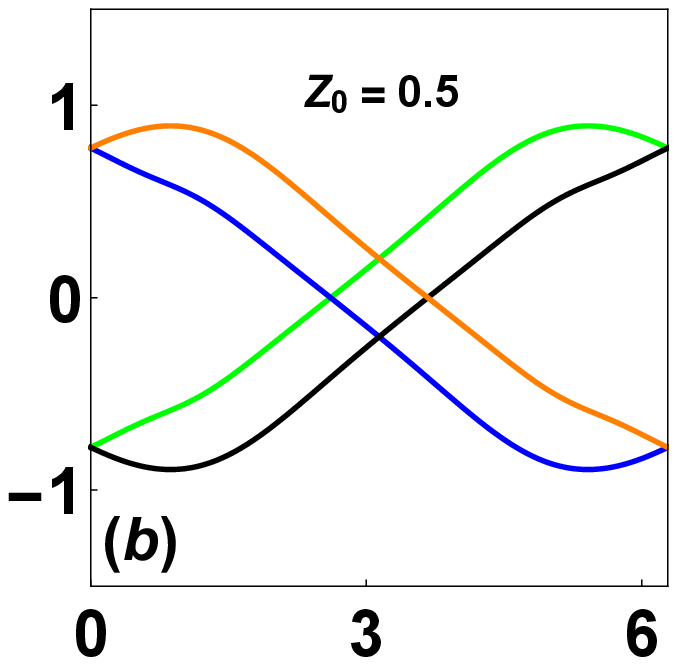}
\hspace{0.01cm}
\vspace{0.1cm}
\includegraphics[scale = 0.47]{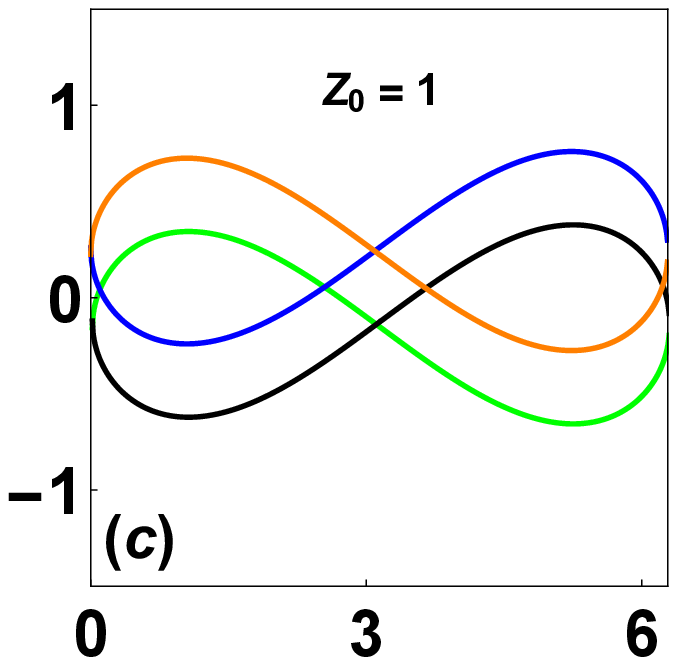}
\hspace{0.01cm}
\vspace{0.1cm}
\includegraphics[scale = 0.47]{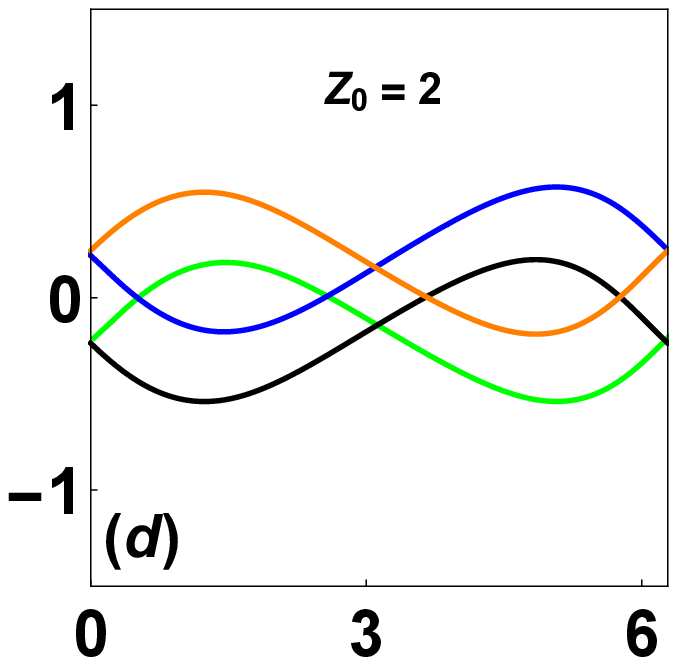}
\hspace{0.01cm}
\vspace{0.1cm}
\includegraphics[scale = 0.47]{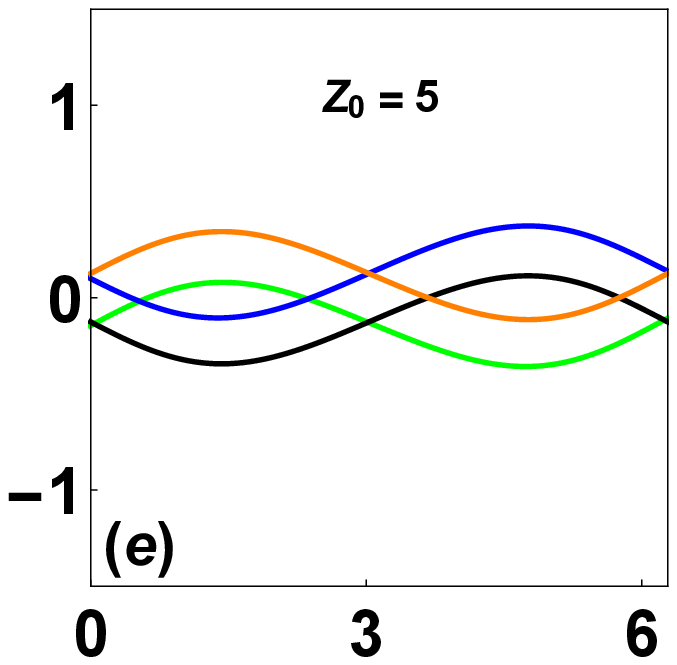}
\hspace{0.01cm}
\vspace{0.1cm}
\includegraphics[scale = 0.47]{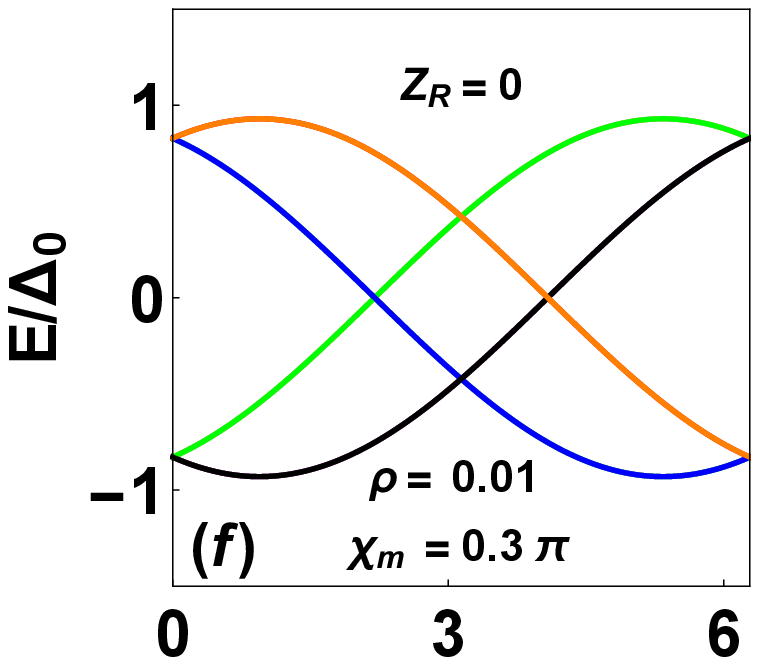}
\hspace{0.01cm}
\vspace{0.1cm}
\includegraphics[scale = 0.47]{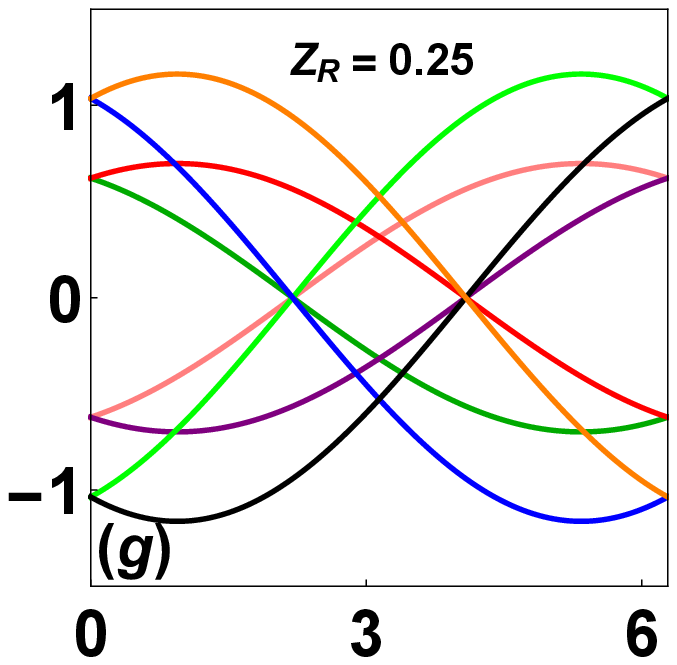}
\hspace{0.01cm}
\vspace{0.1cm}
\includegraphics[scale = 0.47]{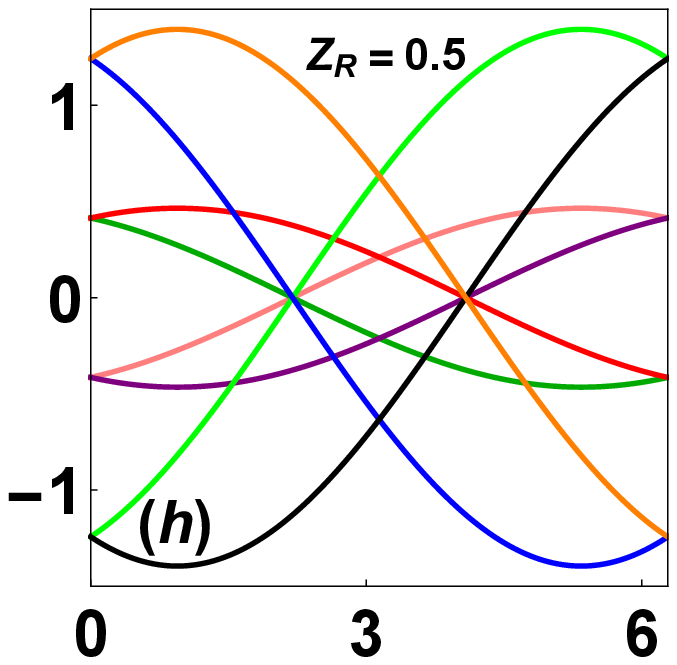}
\hspace{0.01cm}
\vspace{0.1cm}
\includegraphics[scale = 0.47]{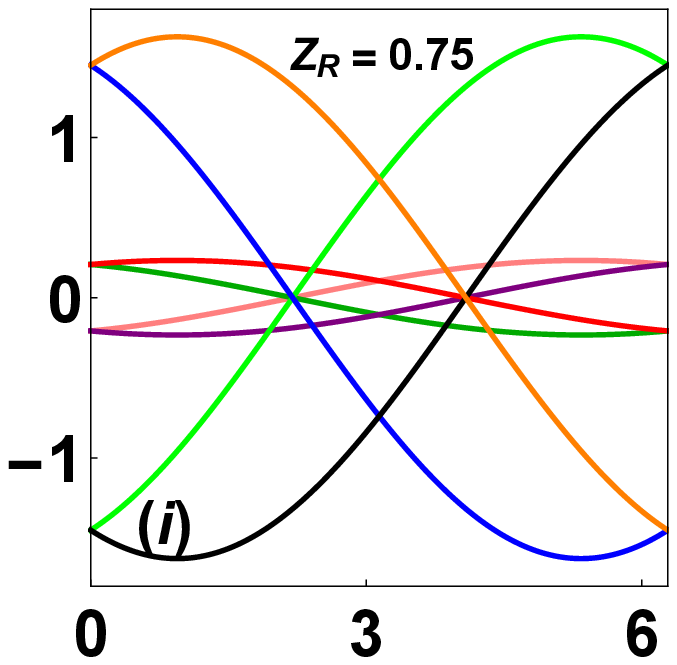}
\hspace{0.01cm}
\vspace{0.1cm}
\includegraphics[scale = 0.47]{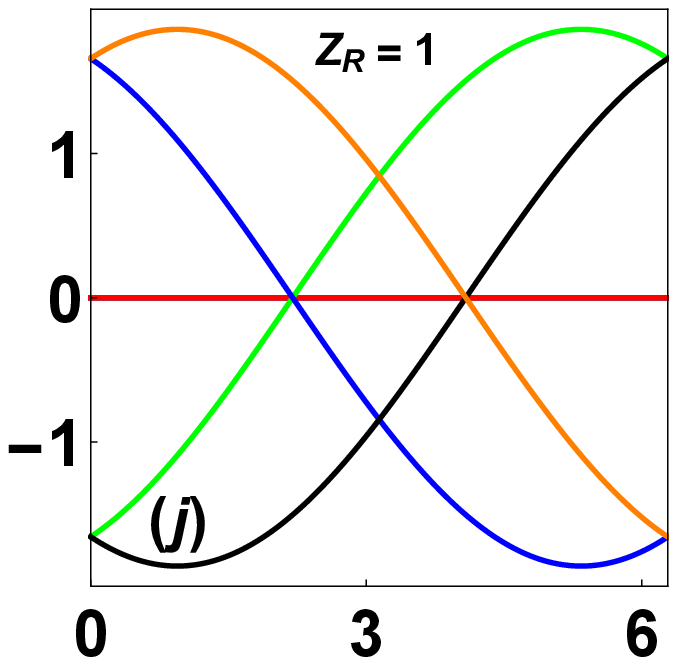}
\hspace{0.01cm}
\includegraphics[scale = 0.47]{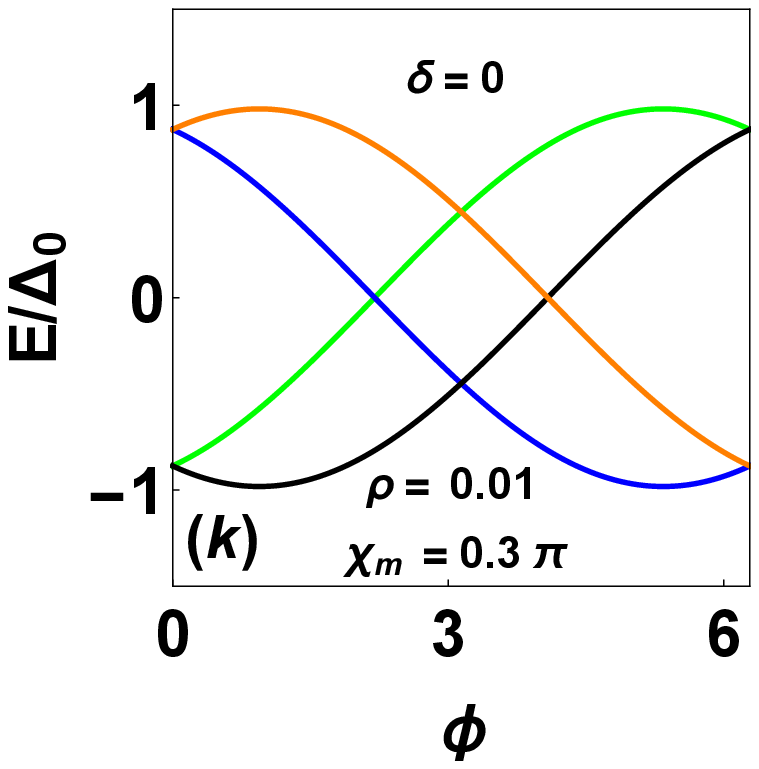}
\hspace{0.01cm}
\vspace{0.1cm}
\includegraphics[scale = 0.47]{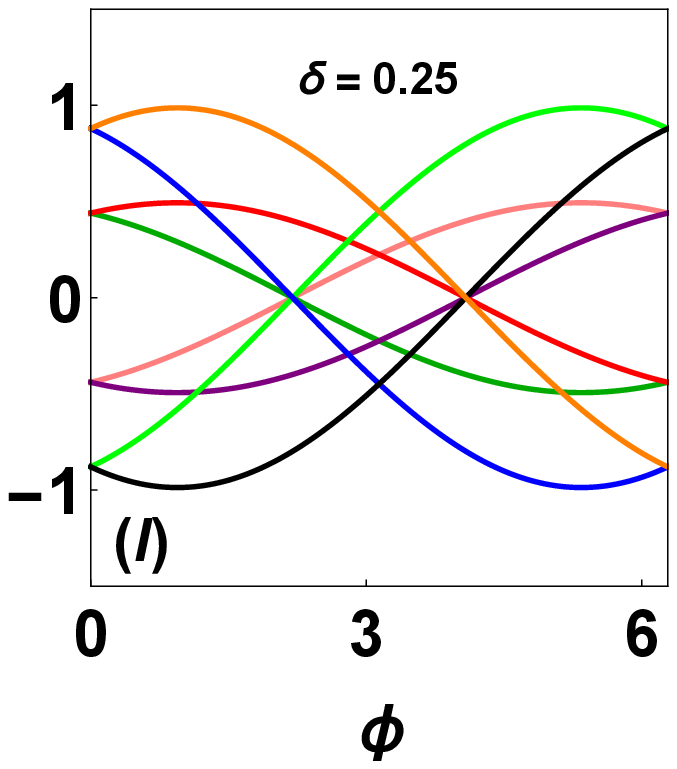}
\hspace{0.01cm}
\vspace{0.1cm}
\includegraphics[scale = 0.47]{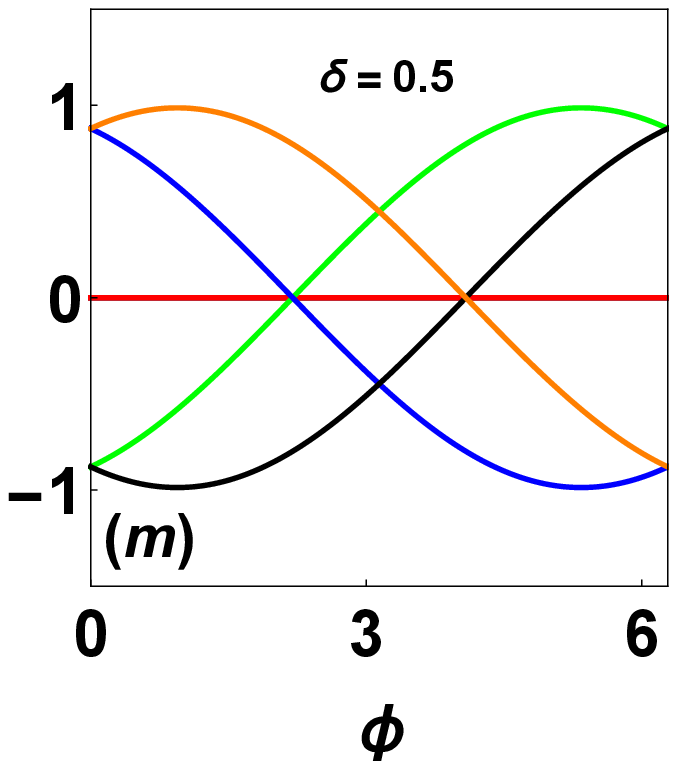}
\hspace{0.01cm}
\vspace{0.1cm}
\includegraphics[scale = 0.47]{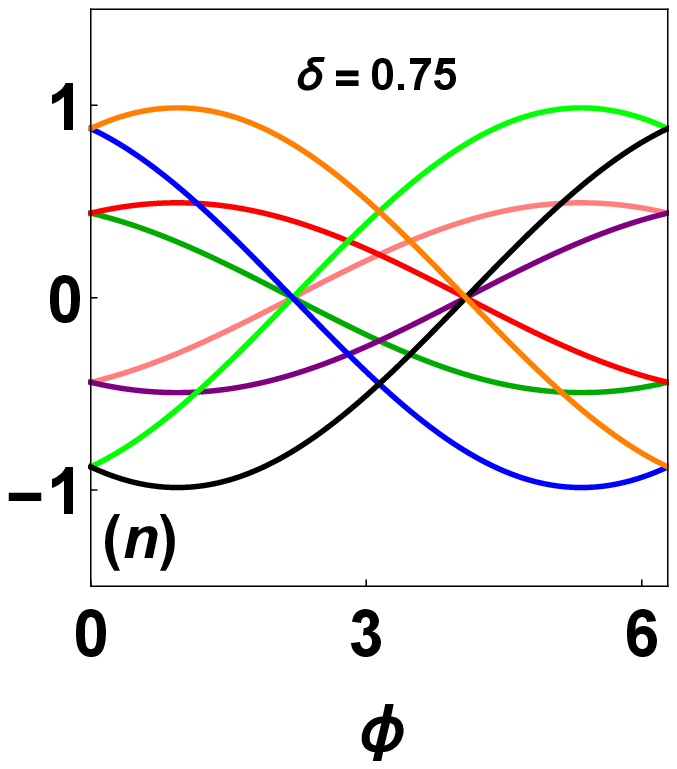}
\hspace{0.01cm}
\vspace{0.1cm}
\includegraphics[scale = 0.47]{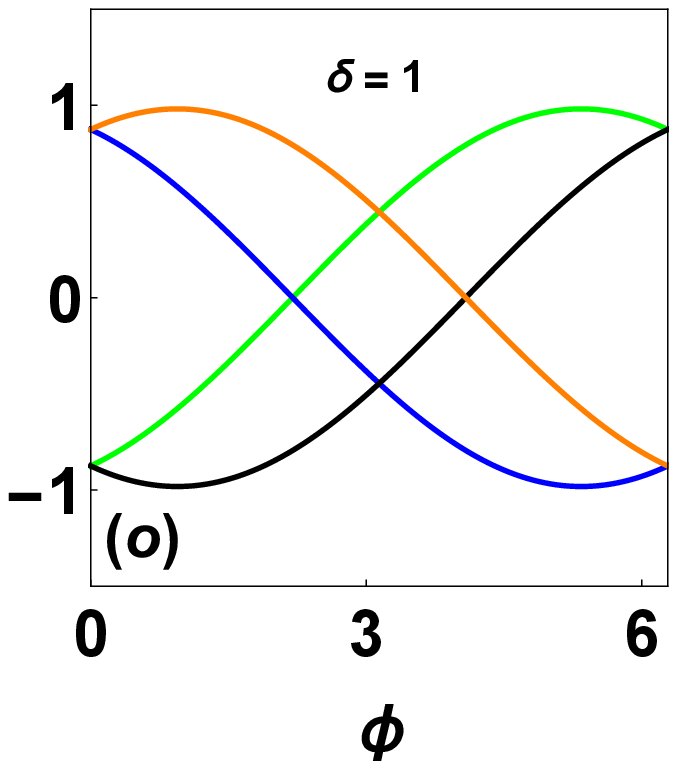}
\hspace{0.01cm}
}
\caption{ABS spectra for different choices of $Z_0$, $Z_R$ and $\delta$. 
We consider $\rho = 0.01$, $\chi_m  = 0.3\pi$, $\xi_m = 0.3\pi$ and 
$L/L_0=0.01$ for all plots. The plots in the top panel are for different 
choices of barrier strength $Z_0$ with $Z_R = 0$ and $\delta = 1$. The plots 
in the middle panel are for different values of RSOC parameter $Z_R$ 
considering $Z_0 = 0$ and $\delta = 1$, while the plots in the bottom panel 
are for different singlet triplet mixing parameter $\delta$ considering 
$Z_0 = 0$ and $Z_R = 0$.}
\label{fig5}
\end{figure*}

We study the effect of barrier transparency on ABS as shown in 
Figs.~\ref{fig5}(a)-\ref{fig5}(e) considering a $\delta$-like barrier at 
$x=0$ and $x=L$. We make this analysis by considering a quasi-magnetic barrier 
with $\rho = 0.01$ and the barrier moment is misaligned by an angle 
$\chi_m = \xi_m = 0.3\pi$ for all plots of Fig.~\ref{fig5}. We also set 
$L/L_0 = 0.01$, $Z_R = 0$ and $\delta = 1$. For a nearly transparent barrier 
the Andreev levels are found to be nearly similar to that obtained in 
Figs.~\ref{fig4}(b), \ref{fig4}(g), \ref{fig5}(a) and \ref{fig5}(b). In this 
scenario, the ABS depart again due to non-vanishing values of $\chi_m$. 
For $Z_0 = 0.5$, the Andreev levels shrinks but the splitting is still present 
as seen in Fig.~\ref{fig5}(b). With the further increase in barrier 
transparency $Z_0$, the energy of the ABS significantly decreases as seen from 
the Figs.~\ref{fig5}(c)-\ref{fig5}(e). It is seen that the pattern of ABS is 
found to be significantly different in Figs.~\ref{fig5}(a) and 
Fig.~\ref{fig5}(e) characterize by an anticrossing at $\phi=0$ and is due to 
finite value of $Z_0$. However, the splitting of the Andreev levels still 
present in the system.

The plots in the middle panel of Fig.~\ref{fig5} display the effect of RSOC on 
Andreev energy levels. In the presence of RSOC, the four degenerate energy 
bands of the TI further split into four branches, as seen in 
Figs.~\ref{fig5}(f)-\ref{fig5}(j). In this situation, the wave vectors of the  
NCSC are different from that of TI. Due to the mismatch of the wave vector, 
an additional gap appears at $\phi = \pi$. However, it is seen from 
Figs.~\ref{fig5}(g)-\ref{fig5}(i) that the four energy branches grow while 
the other four shrinks simultaneously with an increase in the strength of 
RSOC. For $Z_R = 1$, four Andreev energy branches grow to their maximum value 
while the other four branches shrink completely, as seen in 
Fig.~\ref{fig5}(j). In this scenario, thus the supercurrent is carried by 
only one branch of Andreev energy levels. Furthermore, it is noted that though 
the RSOC splits the energy bands, it does not create a phase shift. It 
indicates that there exist Majorana modes at $\phi=\pi$ for normal incidence 
even in the presence of RSOC. Thus, the main effect of RSOC on the Andreev 
levels are, (i) shift of energy bands and (ii) presence of an additional gap 
at $\phi = \pi$. So, it can be concluded that RSOC plays a very significant 
role on the ABS spectrum.

We show ABS spectra for different values of singlet-triplet mixing parameter 
$\delta$ in the bottom panel of Fig.~\ref{fig5}. We set $L/L_0 = 0.01$, 
$Z_R = 0$ and $Z_0 = 0$ for this analysis.  The mixing parameter 
$\delta$ plays a very significant role in NCSC with $\delta=0$ correspond to 
majority of singlet and $\delta=1$ correspond to majority of triplet 
correlations. We find that the system exhibit exactly same behaviour for 
$\delta = 0$ and $1$ as seen from Figs.~\ref{fig5}(k) and \ref{fig5}(o). We 
encounter band splitting again for unequal mixing of the singlet-triplet 
components. For unequal mixing, the Andreev levels corresponding to 
$\Delta_-$ shrinks while the levels corresponding to $\Delta_+$ remains 
unchanged as seen from Fig.~\ref{fig5}(l) and \ref{fig5}(n). Moreover, the 
helical ABS spectra for $\delta = 0.25$ is found to be exactly same as that 
of $\delta = 0.75$. For equal mixing, the Andreev energy levels corresponding 
to $\Delta_-$ die out completely while the levels corresponding to $\Delta_+$ 
remains unchanged as seen from Fig.~\ref{fig5}(m). Thus the supercurrent is 
only due to the $\Delta_+$ component, while the contribution of the 
supercurrent from $\Delta_-$ is zero for equal mixing conditions. We find that 
the magnitude of ABS energy is dependent on gap $\Delta_-$ while it is totally 
independent of $\Delta_+$. Furthermore, the crossing of the energy levels and 
the phase are totally independent of the gap parameter $\Delta_\pm$. It 
signifies that Majorana modes can exist at the surface of 3D TI in a 
$\pi$-NCSC Josephson junction.

\subsection{Josephson Supercurrent}
\begin{figure*}[hbt]
\centerline
\centerline{
\includegraphics[scale = 0.47]{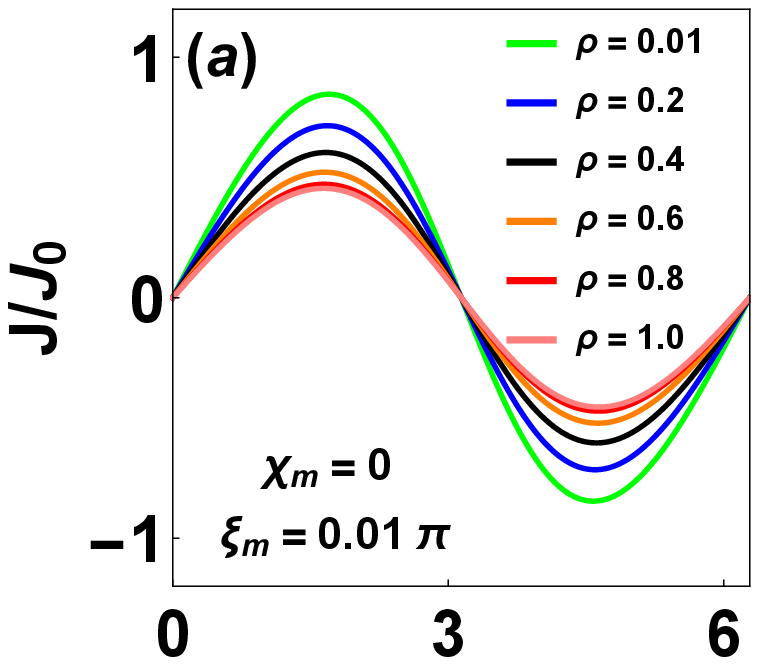}
\hspace{0.001mm}
\vspace{0.1cm}
\includegraphics[scale = 0.47]{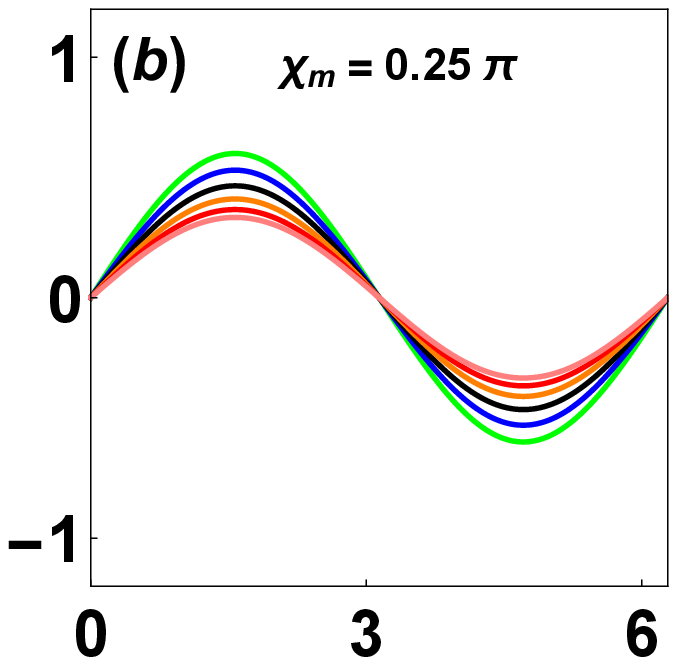}
\hspace{0.001mm}
\vspace{0.1cm}
\includegraphics[scale = 0.47]{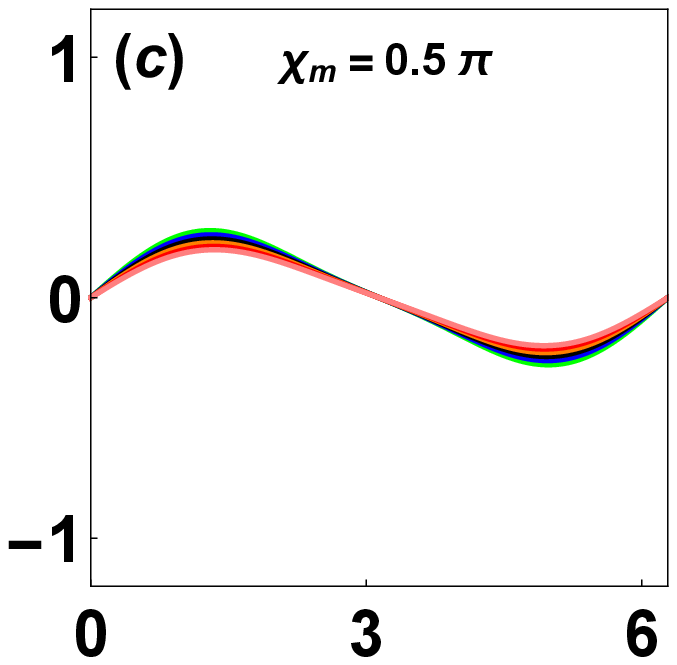}
\hspace{0.001mm}
\vspace{0.1cm}
\includegraphics[scale = 0.47]{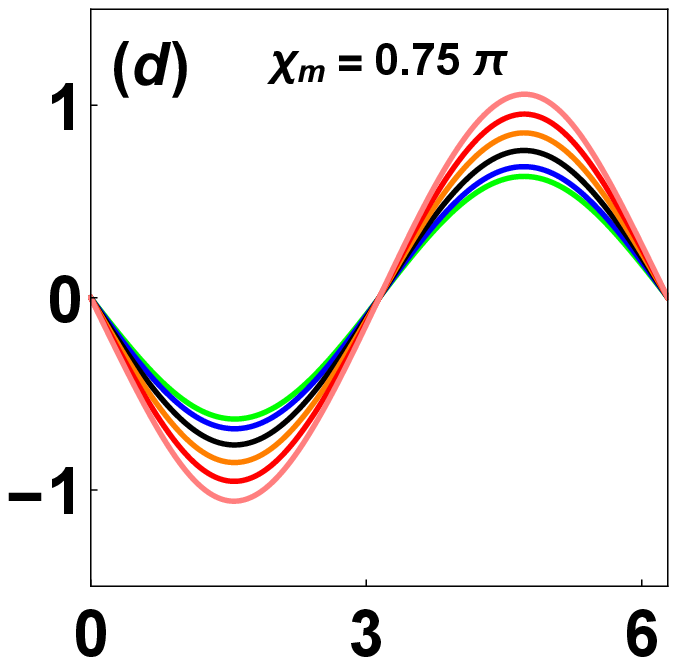}
}
\hspace{0.001mm}
\vspace{0.1cm}
\includegraphics[scale = 0.47]{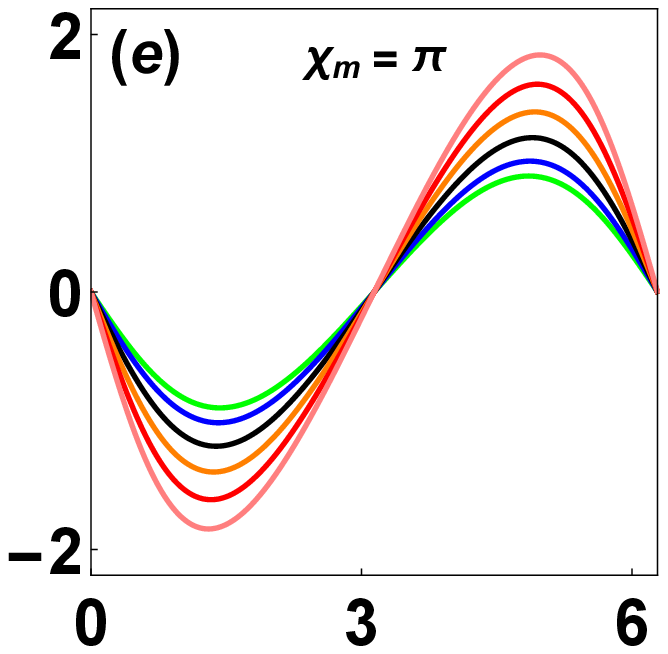}
\hspace{0.001mm}

\includegraphics[scale = 0.47]{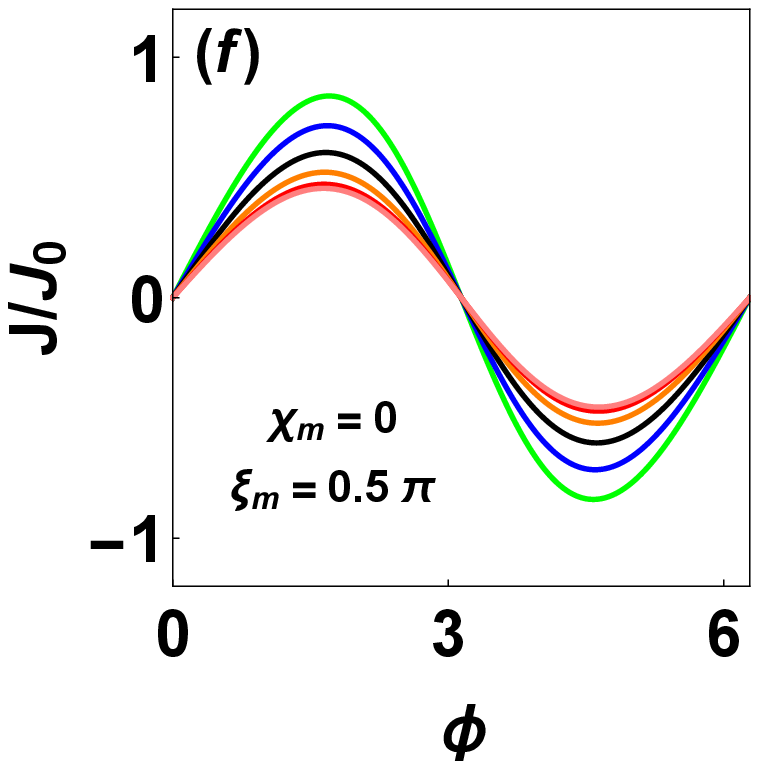}
\hspace{0.001mm}
\vspace{0.1cm}
\includegraphics[scale = 0.47]{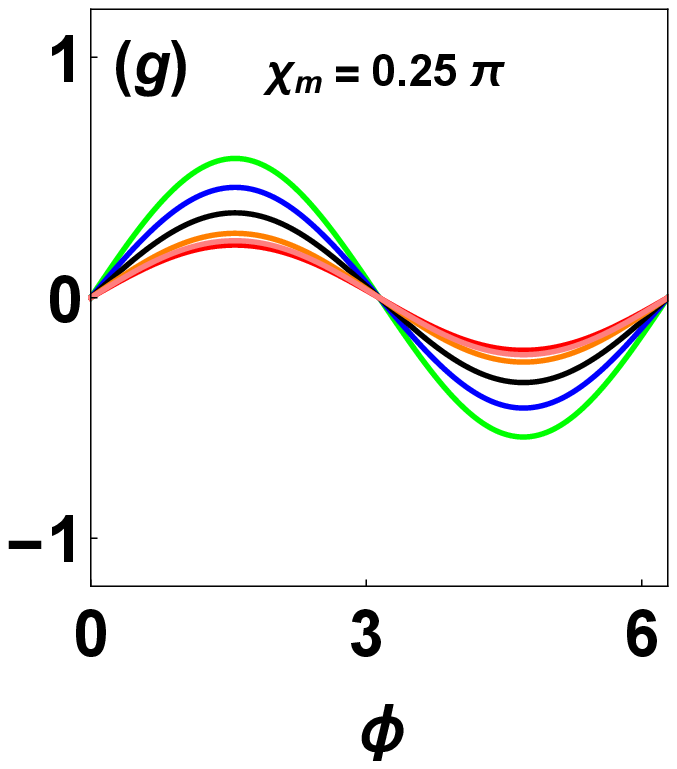}
\hspace{0.001mm}
\vspace{0.1cm}
\includegraphics[scale = 0.47]{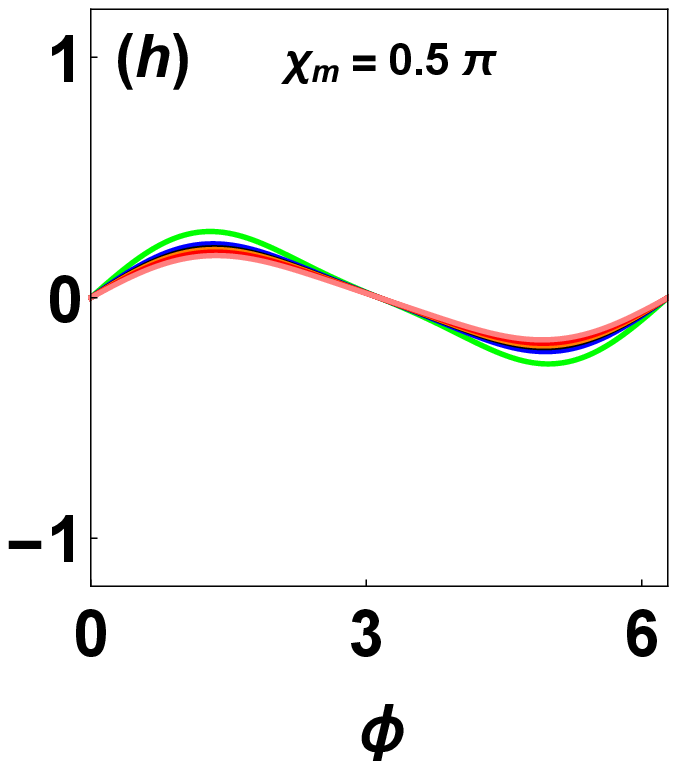}
\hspace{0.001mm}
\vspace{0.1cm}
\includegraphics[scale = 0.47]{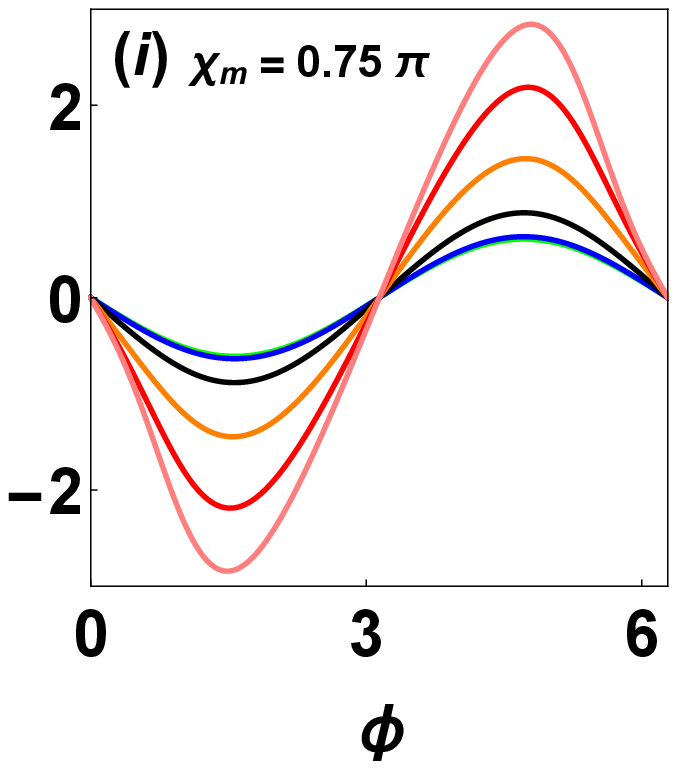}
\hspace{0.001mm}
\vspace{0.1cm}
\includegraphics[scale = 0.47]{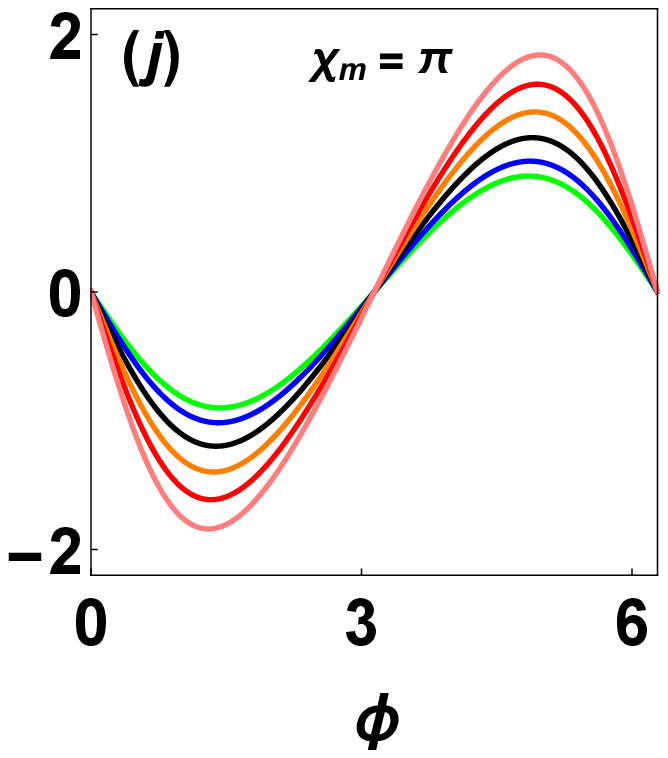}
\hspace{0.001mm}
\caption{Variation of Josephson supercurrent with $\phi$ for different choices 
of $\rho$ and $\chi_m$. The plots in the top panel are for $\xi_m = 0.01\pi$ 
while the plots in the bottom panel are for $\xi_m =0.5\pi$. All the plots are 
for $Z_0 = 0$, $Z_R = 0$ and $\delta =0$ with $L/L_0 = 0.01$.}
\label{fig6}
\end{figure*}

\begin{figure}[hbt]
\centerline
\centerline{
\includegraphics[scale = 0.5]{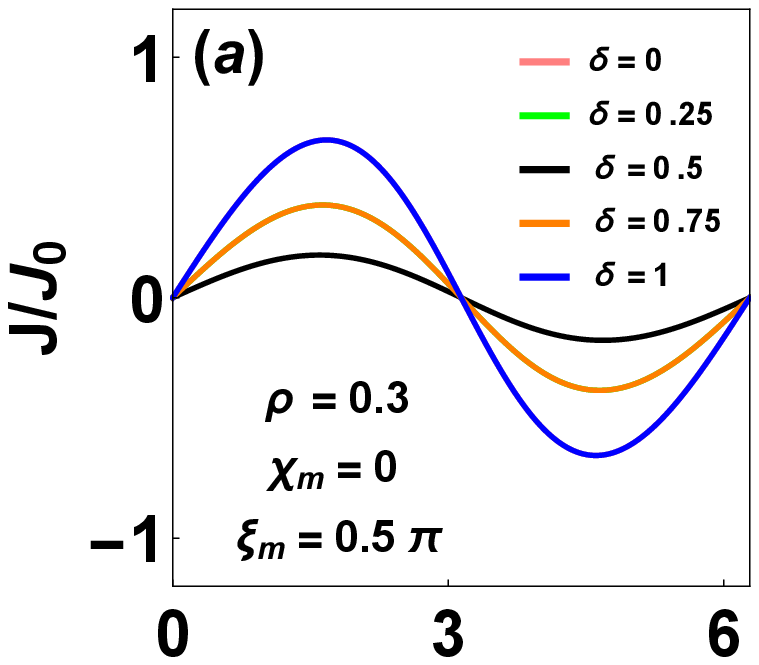}
\hspace{0.001mm}
\vspace{0.1cm}
\includegraphics[scale = 0.5]{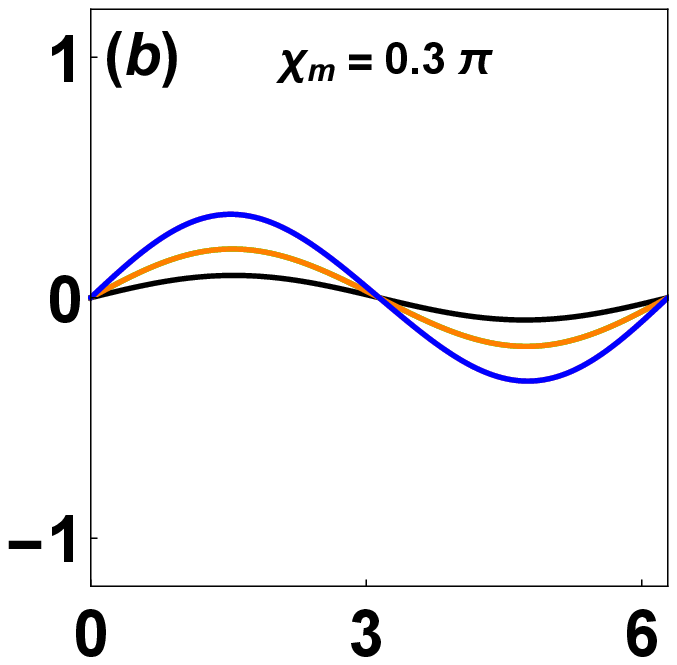}
\hspace{0.001mm}
\vspace{0.1cm}
\includegraphics[scale = 0.5]{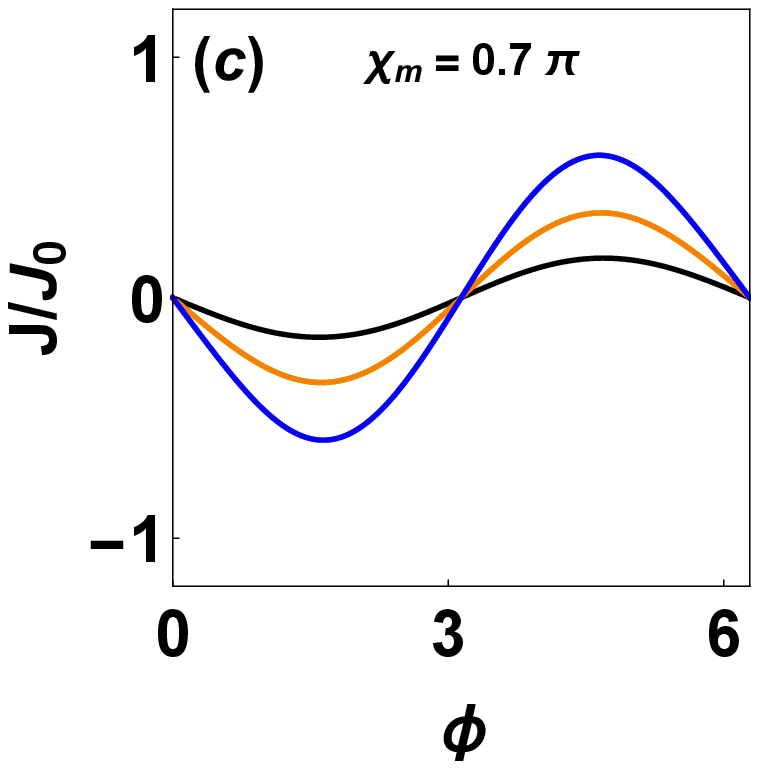}
\hspace{0.001mm}
\vspace{0.1cm}
\includegraphics[scale = 0.5]{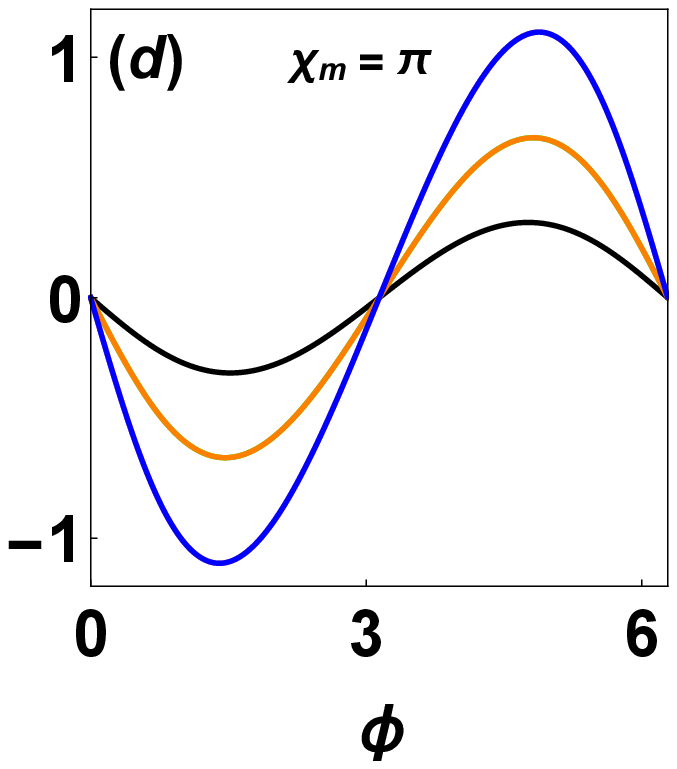}
\hspace{0.001mm}
}
\caption{Variation of Josephson supercurrent with $\phi$ for different choices 
of $\delta$ $\chi_m$ considering $\rho = 0.3$, $\xi_m = 0.5\pi$, $Z_0 = 0$, 
$Z_R = 0$ with $L/L_0 = 0.01$.
}
\label{fig7}
\end{figure}

\begin{figure}[hbt]
\centerline
\centerline{
\includegraphics[scale = 0.5]{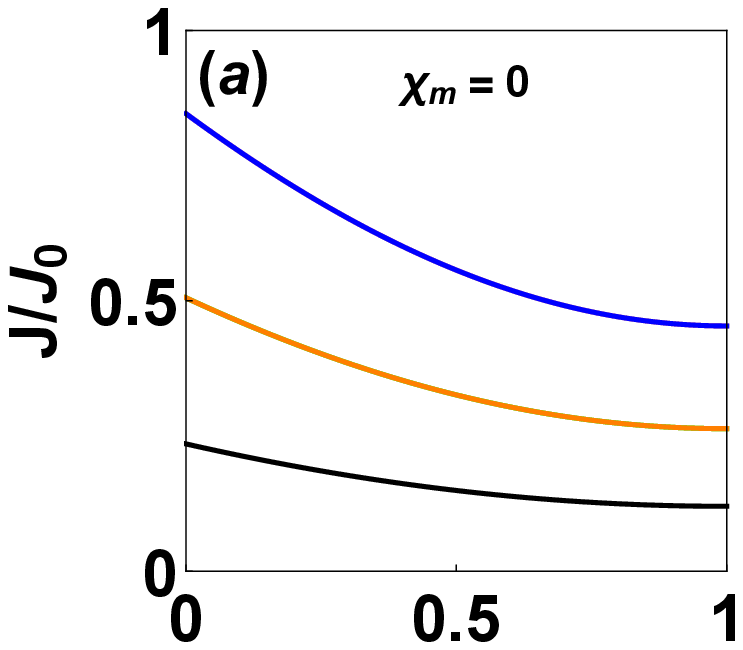}
\hspace{0.001mm}
\vspace{0.1cm}
\includegraphics[scale = 0.5]{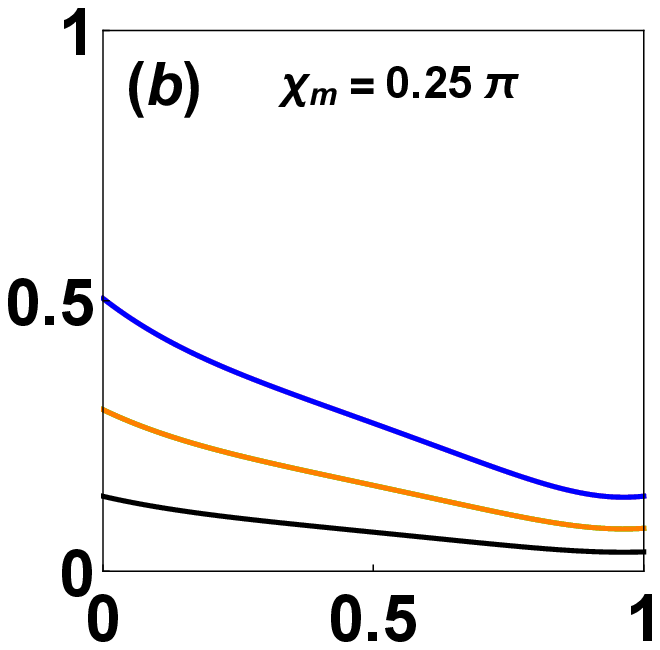}
\hspace{0.001mm}
\vspace{0.1cm}
\includegraphics[scale = 0.5]{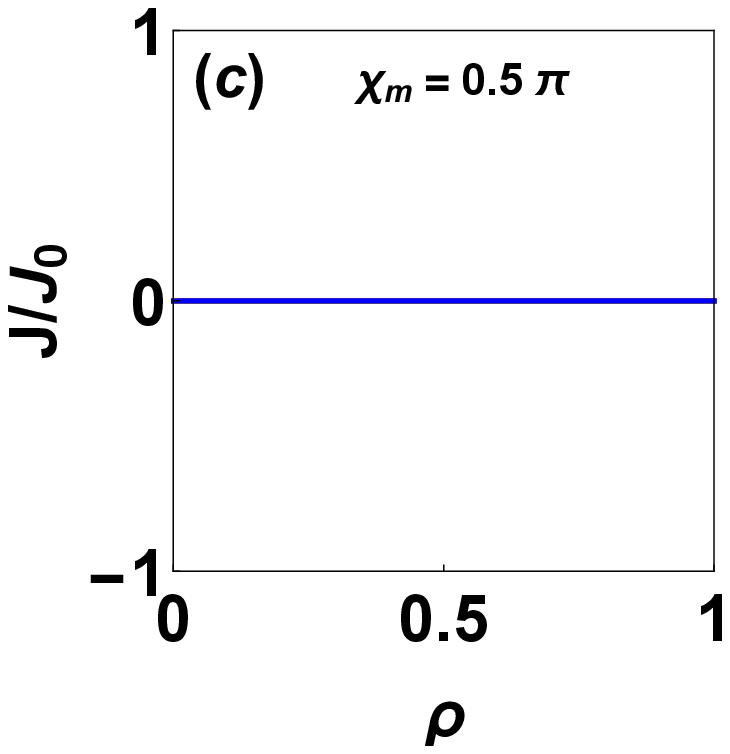}
\hspace{0.01mm}
\vspace{0.1cm}
\includegraphics[scale = 0.5]{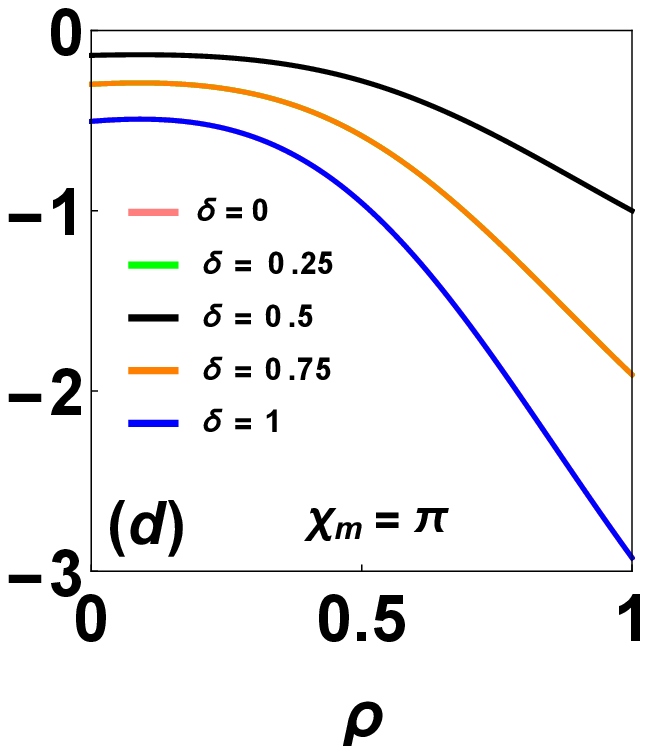}
}
\caption{Variation of Josephson supercurrent with $\rho$ for different choices 
of $\chi_m$ considering $\xi_m = 0.1\pi$, $Z_0 = 0$, $Z_R = 0$ with 
$L/L_0 = 0.01$.}
\label{fig8}
\end{figure}

\begin{figure}[hbt]
\centerline
\centerline{
\includegraphics[scale = 0.39]{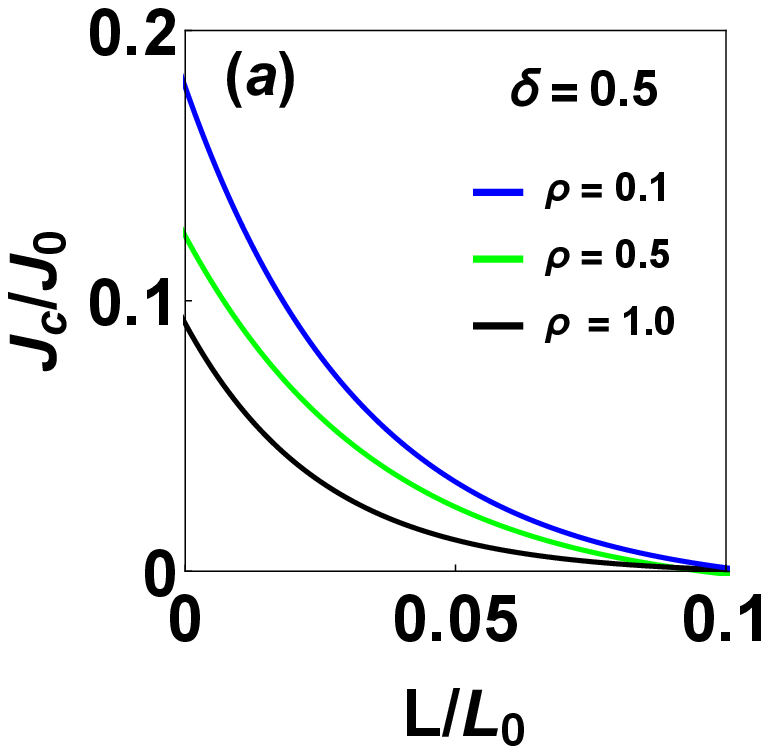}
\vspace{0.1cm}
\includegraphics[scale = 0.39]{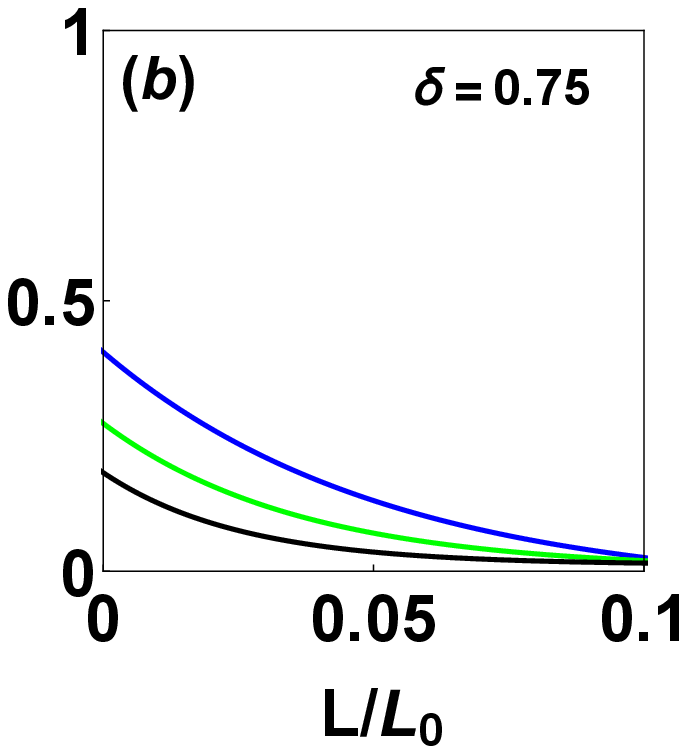}
\includegraphics[scale = 0.39]{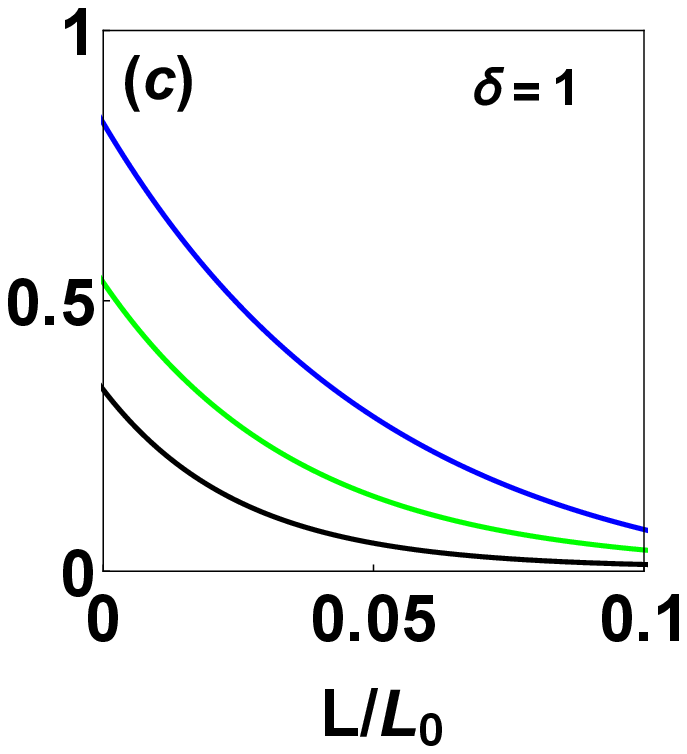}
\hspace{0.001mm}
}
\caption{Plot of length dependence of critical current for different choices 
of $\rho$ with $\chi_m = 0.01\pi$, $\xi_m = 0.5\pi$, $Z_0 = 0$ and $Z_R = 0$.}
\label{fig9}
\end{figure}

\begin{figure*}[hbt]
\centerline
\centerline{
\includegraphics[scale = 0.35]{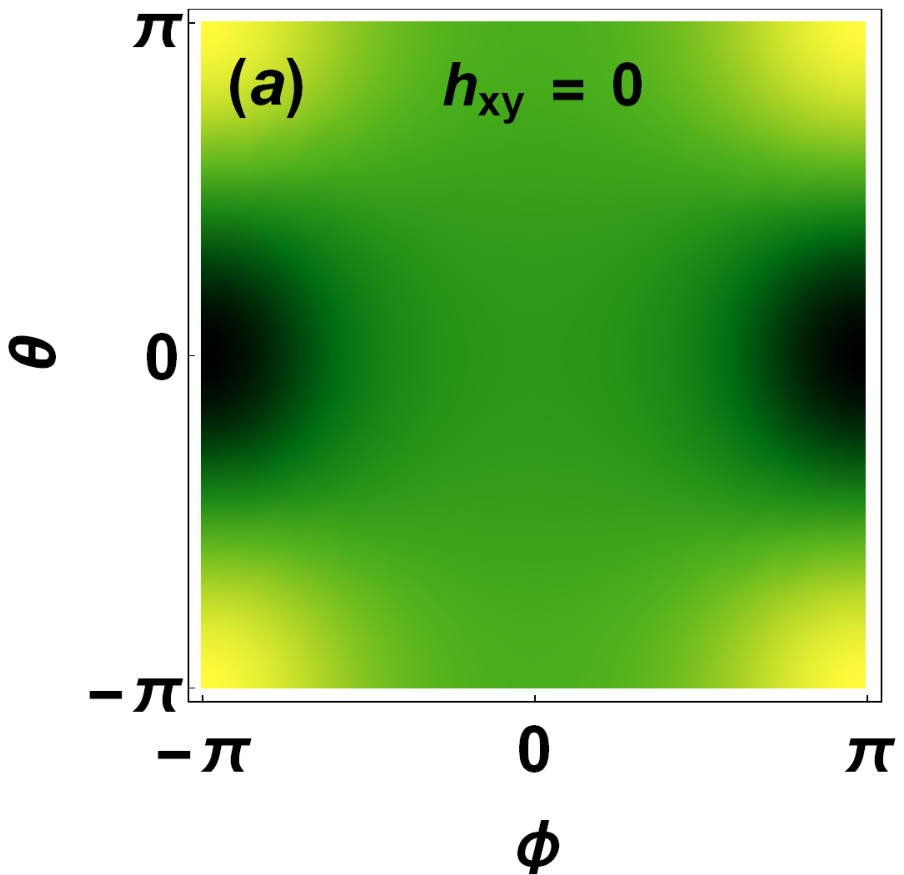}
\hspace{-0.15cm}
\includegraphics[scale = 0.35]{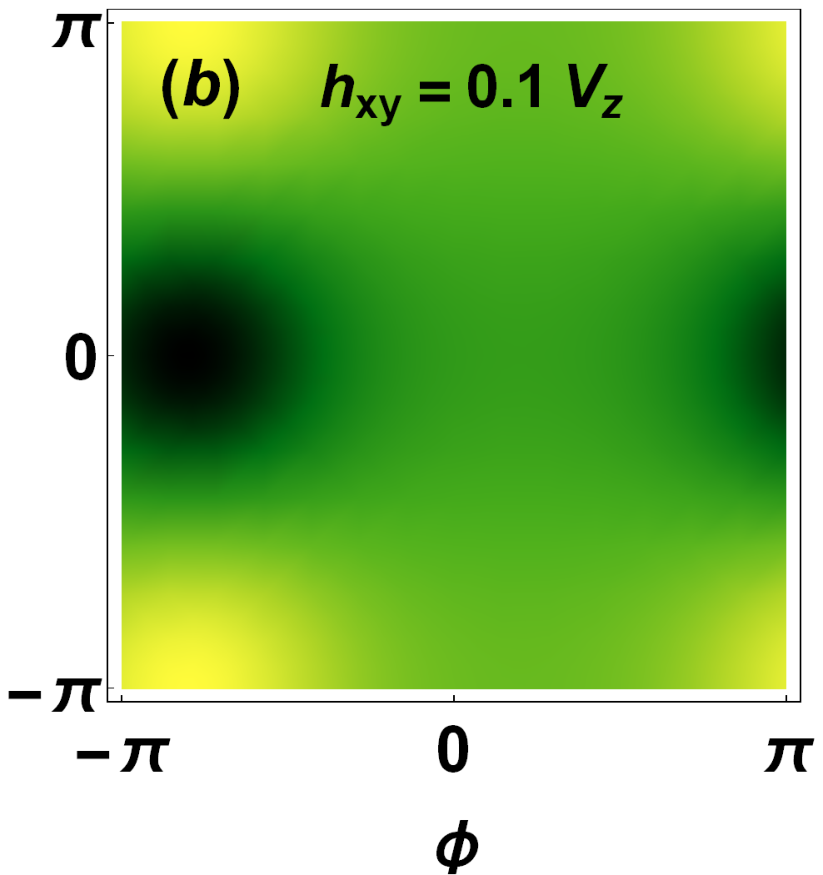}
\hspace{0.05cm}
\includegraphics[scale = 0.337]{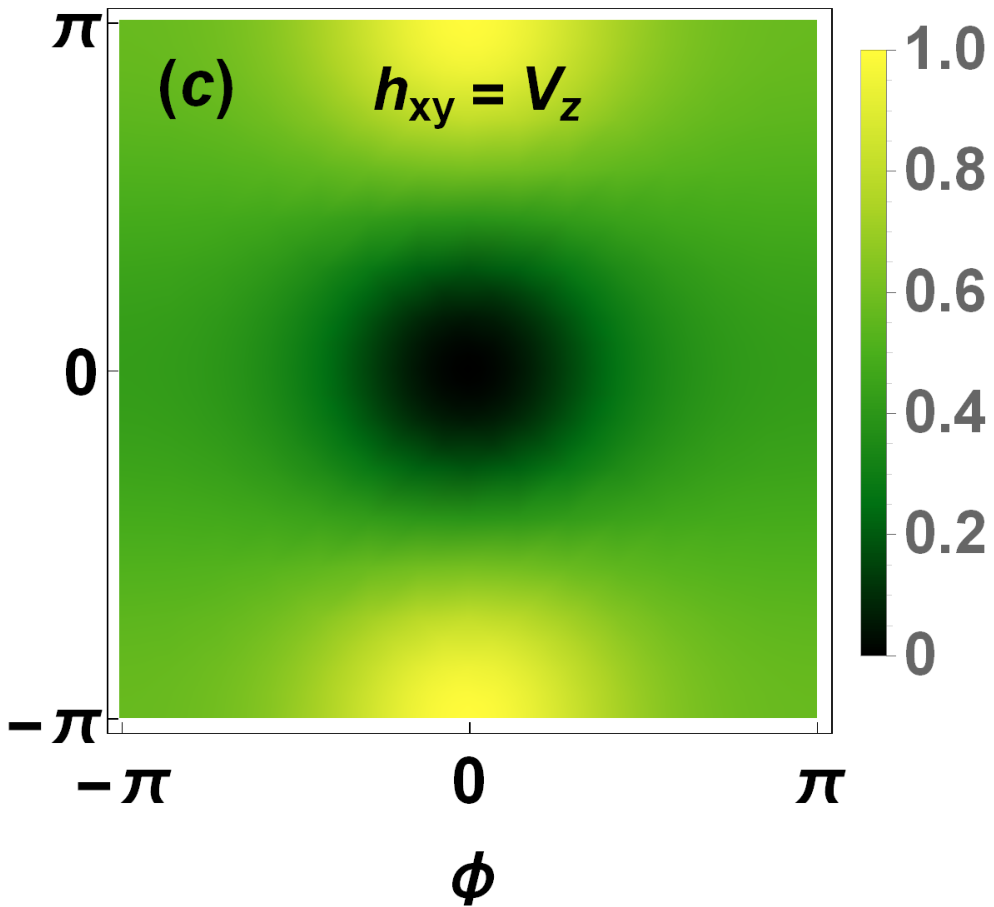}
\includegraphics[scale = 0.4]{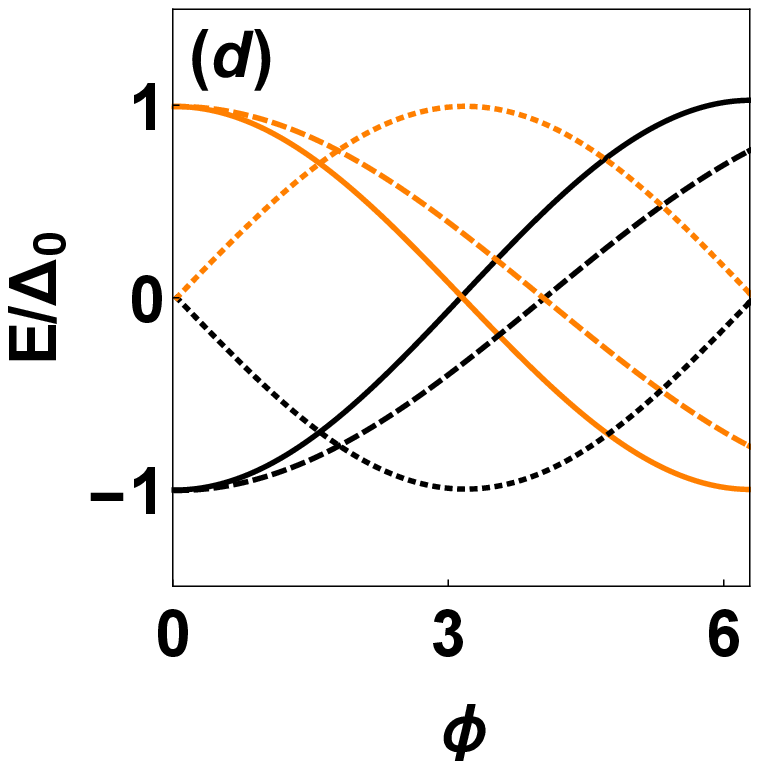}
\includegraphics[scale = 0.4]{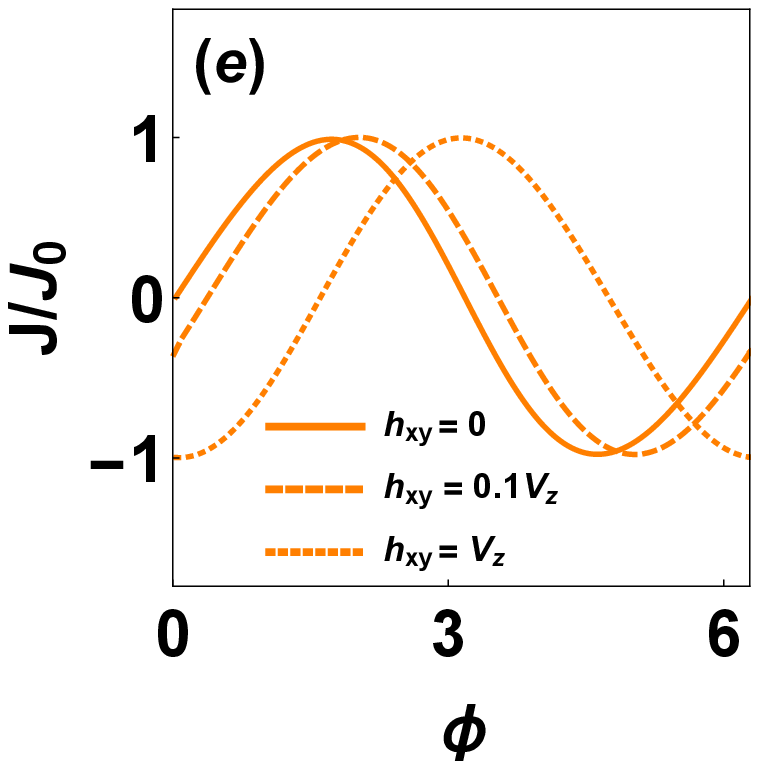}
}
\caption{ABS energy levels $E$ as a function of incident angle $\theta$ and 
phase $\phi$. The plot (a) corresponds to no barrier and bulk magnetization 
i.e., $h_{xy} = 0$, $h_z = V_z = 0$ while plot (b) is for $h_{xy} = 0.1V_z$ 
and plot (c) is for $h_{xy} = V_z$. The corresponding ABS spectra and 
Josephson supercurrent are shown in plots (d) and (e) respectively.}
\label{fig10}
\end{figure*}

The magnetic tunability of the Andreev energy levels and its effects on the 
superconducting phase indicates the presence of a non-trivial Josephson 
current in the proposed setup. Experimentally, Josephson junctions made up of 
Superconductor-Topological Insulator-Superconductor (S$|$TI$|$S) has been 
realized recently \cite{veldhorst}. The work indicates the anomalous 
characteristics of the supercurrent. Moreover, presently there is a great 
interest in the experimental realization of ferromagnetic topological 
insulator hybrids (FTI). It can be achieved by random doping of transition 
metal elements like Cr or V on the surface of 3D TI \cite{chang11,chang12}. 
This result in magnetic proximity on the surface of TI. Recently it is found 
that due to the presence of magnetic material in the NCSC Josephson junction, 
the system displays anomalous Josephson current \cite{zhang11}. It arises due 
to the different orientation of the exchange field. However, the Josephson 
current in an NCSC corresponding to an HM ferromagnet is not studied yet. 
Also, as NCSC$|$HM$|$NCSC Josephson junctions hold Majorana modes at 
$\phi=\pi$ in parallel orientation for half-metallic limit, so it is necessary 
to understand the behaviour of supercurrent flowing at the surface of 3D TI. 
We initially calculated the Josephson supercurrent for an HM with bulk magnetic 
moments $\mathbf{m} = (0,0,|m|)$ and the barrier magnetic moments 
$\mathbf{V}_m=(V_x,V_y,V_z)=(\rho V_0 \cos\xi_m\sin\chi_m, \rho V_0 \sin\xi_m\sin\chi_m, \rho V_0 
\cos\chi_m)$. The results in the half-metallic limit are displayed in 
Figs.~\ref{fig6}, \ref{fig7}, \ref{fig8} and \ref{fig9}. However, we also 
plotted the supercurrent for a general ferromagnet in Fig.~\ref{fig10}. For 
simplicity of our calculations, we ignore the continuum states of asymmetric 
contact junction and we only focus on the discrete states. This approximation 
does not drastically change the effective current phase relation and hold good 
to simplify the calculations as seen from the previous works \cite{klam}.

In Fig.~\ref{fig6}, we have plotted the Josephson supercurrent for different 
values of the effective ratio of magnetic and non-magnetic barrier potentials. 
We find that for parallel orientation of the barrier and bulk moments with 
$\xi_m = 0.01\pi$, the magnitude of supercurrent is maximum for $\rho = 0.01$. 
It decreases with the increasing values of $\rho$ as seen from 
Fig.~\ref{fig6}(a). This signifies that in this situation, a non-magnetic 
barrier offers more supercurrent than the magnetic barrier. It is also noted 
that the supercurrent is found to be 2$\pi$ periodic. If the barrier and the 
bulk moments are misaligned to an angle $\chi_m = 0.25\pi$, then the 
supercurrent significantly reduces for all choices of $\rho$. However, in this 
condition also non-magnetic barrier offer more supercurrent than the magnetic 
barrier and the supercurrents still found to display $2\pi$ periodic phase as
seen from Fig.~\ref{fig6}(b). This is due to the presence of symmetric 
branches in ABS spectra in HM limit, which is significantly different for a 
bulk ferromagnet \cite{zhang11}. The magnitude of the supercurrent further 
decreases for the perpendicular misalignment of the barrier and bulk moments 
characterize by $\chi_m = 0.5\pi$. In this case, all the magnetic barrier 
corresponding to non-zero values of $\rho$ display exactly the same 
characteristics. As the misalignment angle $\chi_m$ increases further to 
$0.75\pi$ and $\pi$, the supercurrents display $\pi$ phase shift. But in this 
case, the component $\rho = 1$ display critical current while it is found to be minimum for $\rho=0.01$. Thus the magnetic barriers enhance the supercurrent 
while the non-magnetic barriers suppress the supercurrent, as seen from 
Figs.~\ref{fig6}(d) and \ref{fig6}(e). It is to be noted that the magnitude of 
the supercurrent is found to be maximum for the anti-parallel misalignment of 
the barrier and the bulk moments. To understand the effect of azimuthal angle 
on the supercurrent, we plotted the CPR for $\xi_m = 0.5\pi$ in 
Figs.~\ref{fig6}(f)-\ref{fig6}(j). It is seen that the nature of the 
supercurrent remains quite similar for the $\chi_m=0$ case as seen from 
Fig.~\ref{fig6}(f). However, for $\chi_m=0.25\pi$ we find that the supercurrent
corresponding to $\rho = 0.01$ and $0.2$ enhances while it suppressed  further 
for $\rho = 0.8$ and $0.1$ as seen from Fig.~\ref{fig6}(g). Supercurrent is 
found to be independent of the strength of the magnetic barrier for 
$\chi_m=0.5\pi$. However, it is still found to be maximum for non-magnetic 
barrier that is seen from Fig.~\ref{fig6}(h). The CPR for $\chi_m=0.75\pi$ 
from Fig.~\ref{fig6}(i) is found to drastically different from 
Fig.~\ref{fig6}(d). In this condition, the magnitude of the supercurrent 
corresponding to the magnetic barrier drastically increases while it 
suppressed for a non-magnetic barrier. It indicates that CPR is also dependent 
on the azimuthal angle of the barrier moment. For the anti-parallel 
misalignment, we find exactly similar characteristics from Figs.~\ref{fig6}(e) 
and \ref{fig6}(j). Thus these results indicate that supercurrent is dependent 
on the misalignment of barrier and bulk magnetic moments. Depending upon the 
orientations, both magnetic and non-magnetic barriers offer critical current 
in certain orientations.

We study the CPR of the total supercurrent for different singlet-triplet 
correlations in Fig.~\ref{fig7}. We set $\rho = 0.3$, $\xi_m = 0.5\pi$, 
$Z_0 = 0$, $Z_R = 0$ and $L/L_0 = 0.0$1 for this analysis. We find that the 
CPR corresponding to $\delta = 0$ coinciding with that for $\delta = 1$ which 
is in accordance with Fig.~\ref{fig5}. It is to be noted that the critical 
current is observed for the majority of either singlet or triplet components 
represented by $\delta = 0$ and $1$ respectively. The critical current is 
found to be minimum for equal mixing of the singlet-triplet parameter for 
all possible misalignment angles. It is due to the reason that for equal 
mixing, the supercurrent is only due to the $\Delta_+$ component while it 
is completely suppressed for the $\Delta_-$ component represented by solid 
black lines of Fig.~\ref{fig7}. However, the critical current display a 
moderate value for unequal mixing of the singlet-triplet parameter 
represented by solid orange lines of Fig.~\ref{fig7}. In this condition, 
though the ABS is suppressed due to the $\Delta_-$ component, some 
supercurrent still presents due to the non-vanishing value of $\Delta_-$. 
Moreover, the CPR is found to execute nearly sinusoidal characteristics as 
the contribution from the higher harmonics are very much suppressed.


From Figs.~\ref{fig6} and \ref{fig7}, we have seen that the characteristics of 
Josephson supercurrent are dependent on the effective ratio between magnetic 
and non-magnetic moments, misalignment angle between the barrier and bulk 
magnetic moment and singlet-triplet mixing parameter. So to understand the 
interplay between them, we have plotted supercurrent as a function of $\rho$ 
in Fig.~\ref{fig8} for different singlet-triplet mixing parameter. We find 
that for $\chi_m=0$, the critical current is observed at $\rho = 0$ and the 
supercurrent decreases monotonically with the increasing values of $\rho$ for 
all choices of singlet-triplet mixing parameter as seen from 
Fig.~\ref{fig8}(a). The maximum critical current is observed for 
$\delta = 0$ and $1$ while the minimum critical current is observed for 
$\delta = 0.5$ as seen earlier from Figs.~\ref{fig6} and \ref{fig7}. The 
characteristics of the critical current remain the same for $\chi_m = 0.25\pi$ 
as seen from Fig.~\ref{fig8}(b). However, the critical current is found to be 
decreased in all mixing scenarios for $\chi_m = 0.25\pi$. For 
$\chi_m = 0.5\pi$, the Josephson current is found to be totally independent of 
$\delta$ for all $\rho$ values as seen from Fig.~\ref{fig8}(c). We find an 
exactly opposite behaviour characterized by the negative critical current for 
anti-parallel misalignment. In this case also $\delta = 0$ and $\delta = 1$ 
display maximum critical current as seen from Fig.~\ref{fig8}(d). 

We study the dependence of the critical Josephson current on the length of the 
Josephson junction in Fig.~\ref{fig9}. It is a quantity that represents the 
maximum supercurrent and often measured in magnetic Josephson junctions 
\cite{linder12,snelder1,snelder2}. We consider $\xi_m = 0.5\pi$, $Z_0 = 0$ and 
$Z_R = 0$ for this analysis. We find that the critical current monotonically 
decays for all choices of mixing parameter $\delta$. The misalignment angle 
can only split the energy bands but not transfer any finite centre of mass 
momentum to the Cooper pairs. This results in a monotonous decay of the 
critical current. From Fig.~\ref{fig9} we see that $\rho = 0.1$ provides 
the maximum critical currents in all three scenarios, while minimum critical 
current is observed for $\rho = 1$. It is also seen that the magnitude of 
critical current is dependent on the mixing parameter $\delta$. The critical 
current is found on decaying rapidly in an equal mixing scenario as seen from 
Fig.~\ref{fig9}(a). For unequal mixing $\delta = 0.75$ and $1$, the critical 
current decays comparatively slower than  that for $\delta = 0.5$ as seen from 
Figs.~\ref{fig9}(b) and \ref{fig9}(c). The decay rate of critical current is 
found to be minimum for $\delta = 1$. Thus it can be concluded that the decay 
rate of the critical current is dependent on the singlet-triplet mixing 
parameter.

Figs.~\ref{eq4}, \ref{fig5} and \ref{fig6} indicate that the presence of a 
spin-active barrier drastically changes the characteristics of Andreev energy 
levels and Josephson supercurrent in an HM ferromagnet. For all our analysis, 
we consider the magnetization of the bulk HM ferromagnet is along z-direction 
and contribution due to xy-component is zero, i.e., $h_{xy}=0$. However, it is 
natural to address the scenario of a general ferromagnet which can be 
represented by random orientation of the bulk moments. Though many similar 
works \cite{tanaka1,zhang11,linder12} is done before but the interplay of the 
bulk moment with spin-active barrier moments on the surface of a TI is yet to 
be studied. So, we have plotted ABS energy levels as a function of $\theta$ 
and $\phi$ in Fig.~\ref{fig10}. We consider three different bulk moments. In 
Fig.~\ref{fig10}(a), we consider the scenario where no barrier and bulk moment 
is present, i.e., $h_{xy}=0$, $h_z = V_z = 0$. In this scenario, we observe a 
4$\pi$ periodic ABS and a band crossover at $\phi = -\pi$ and $\pi$ specif by 
the darker region of the density plot of Fig.~\ref{fig10}(a). In this condition,
 the Andreev energy levels are found to be symmetric with respect to $\phi = 0$ and Majorana modes are observed at $\phi = \pi$. However, for $h_{xy}=0.1V_z$, 
the band crossing will not appear at $-\pi$ and $\pi$. Thus the ABS spectra 
are found to be asymmetric in this condition seen from Fig.~\ref{fig10}(b). In 
Fig.~\ref{fig10}(c), we display the ABS for $h_{xy}= V_z$. In this condition, 
the crossover of the Andreev energy bands will appear at $\phi = 0$, which 
signify the $2\pi$-periodic ABS. The ABS spectra and CPR for all the choices 
of $h_{xy}$ are shown in Figs.~\ref{fig10}(d) and \ref{fig10}(e). We find that 
there exists a phase shift in Josephson supercurrent for random orientation of 
the bulk magnetic moments.

It is very worthwhile to mention that experimentally transport properties of 
heterostructures are studied in the diffusive limit. In the diffusive limit, 
one needs to use quasi-classical Usadel equations to obtain the Andreev levels 
and supercurrent. However, an analytic equation for ABS energy considering all 
contributing parameters is too hard to obtain for heterostructure like 
NCSC$|$HM$|$NCSC. So in that case also we need to use a numerical approach. 
Though we perform all our calculations considering BdG equations in the 
ballistic limit, our results are quite adequate in diffusive limits also 
because of two reasons: (1) The anomalous Andreev energy levels and 
the $\phi$-states are generated due to chiral spin symmetry breaking along 
with scattering from the spin-active barrier. It is to be noted that these 
effects pertain uniquely to both ballistic and diffusive limits. (2) We find 
that CPR for low and moderate interface transparency is close to sinusoidal 
in nature. It is due to the suppression of the higher harmonics. As in 
diffusive limit, first harmonics give the most important contribution. So our 
results will be quite adequate in diffusive limits also. There may be some 
difference in ballistic and diffusive limits qualitatively. But it is expected 
to have no dramatic alteration of the results in diffusive limit.

\section{Conclusions}
In summary, we have studied the transport in NCSC$|$HM$|$NCSC Josephson 
junction grown on a 3D TI. Our model predicts the formation of the helical 
Andreev Bound States, Majorana mode and Josephson supercurrent on the surface 
of a 3D TI. We employ Bogoliubov de-Gennes equations to obtain ABS and CPR. 
We investigated the role of the spin-active barrier by considering different 
alignments of barrier magnetic moment on Andreev levels and supercurrent. It is
seen that a 4$\pi$-periodic ABS is formed for parallel alignment of barrier and
bulk magnetic moments for all incidence angles. However, for anti-parallel 
alignment, the system resides in 2$\pi$ in Andreev levels. It is seen that 
Majorana modes exist at $\pi$ for parallel alignment in normal incidence, while
they disappear in oblique incidence condition. We find that the degenerate 
Andreev bands split into their corresponding branches for misalignments of the 
barrier and bulk moments. In this condition, a phase shift is observed, which 
results in $\phi$-junction. It is seen that the Andreev energy levels are 
suppressed for a strongly magnetic barrier in parallel orientation conditions, 
while a reverse scenario is observed for anti-parallel orientation. The 
magnitude of ABS energy decreases with the increase in barrier width. It is 
found that there exists an anti-crossing at $\phi=0$ in the ABS spectra for 
a finite barrier thickness. In the presence of RSOC, the degenerate sub-bands 
further split into their respective branches. It is seen that with the 
increasing strength of RSOC, the Andreev levels corresponding to positive 
spin helicity grow, while the other shrinks. For an NCSC with very strong RSOC,
the ABS branches corresponding to negative spin helicity shrinks completely 
result in no supercurrent for these branches. We find that the band splits for 
unequal mixing of singlet-triplet parameter even in the absence of RSOC. 
However, no splitting is observed for equal mixing or singlet/triplet 
dominated scenarios. We also find that for parallel alignment of the moments, 
a non-magnetic barrier offers more supercurrent than a magnetic barrier. 
However, an opposite scenario is observed in anti-parallel orientation. So 
depending on the orientations, both magnetic and non-magnetic barriers offer
critical current in certain orientations. The decay rate of the critical 
current is found to be dependent on the singlet-triplet mixing parameter. We 
also find an anomalous characteristic of Josephson supercurrent for random 
orientation of the bulk magnetic moment, but it is not observed in the 
half-metallic limit. These results would be a significant guiding background 
for innovating novel TI based sprintronics appliances.

\end{document}